\documentclass[aps,pra,twocolumn,superscriptaddress,nofootinbib,amsmath,amssym,floatfix]{revtex4-2}

\usepackage{enumitem}                  
\usepackage{float}                     
\usepackage{graphicx}                  
\usepackage{graphics}                  
\usepackage{amsmath}                   
\usepackage{commath}                   
\usepackage{bm}
\usepackage{booktabs}                  
\usepackage{multirow}
\usepackage{datetime}                  
\usepackage{hyperref}
\usepackage[export]{adjustbox}         
\usepackage{xcolor}                    
\let\oldhat\hat
\renewcommand{\hat}[1]{\oldhat{\mathbf{#1}}}
\usepackage[normalem]{ulem} 

\begin{document}

\title{Probing the position-dependent optical energy fluence rate in three-dimensional scattering samples}

\author{Ozan Akdemir}
\email{o.akdemir@utwente.nl}
\affiliation{Complex Photonic Systems (COPS), MESA+ Institute for Nanotechnology, 
University of Twente, P.O. Box 217, 7500 AE Enschede, The Netherlands}

\author{Minh Duy Truong}
\affiliation{Complex Photonic Systems (COPS), MESA+ Institute for Nanotechnology, 
University of Twente, P.O. Box 217, 7500 AE Enschede, The Netherlands}

\author{Alfredo Rates}
\affiliation{Complex Photonic Systems (COPS), MESA+ Institute for Nanotechnology, 
University of Twente, P.O. Box 217, 7500 AE Enschede, The Netherlands}

\author{Ad Lagendijk}
\affiliation{Complex Photonic Systems (COPS), MESA+ Institute for Nanotechnology, 
University of Twente, P.O. Box 217, 7500 AE Enschede, The Netherlands}

\author{Willem L. Vos}
\email{w.l.vos@utwente.nl}
\affiliation{Complex Photonic Systems (COPS), MESA+ Institute for Nanotechnology, 
University of Twente, P.O. Box 217, 7500 AE Enschede, The Netherlands}

\date{23 July 2024, for Phys. Rev. A. } 

\begin{abstract}
The accurate determination of the position-dependent energy fluence rate of scattered light (that is proportional to the energy density) is crucial to the understanding of transport in anisotropically scattering and absorbing samples, such as biological tissue, seawater, atmospheric turbulent layers, and light-emitting diodes.
While Monte Carlo simulations are precise, their long computation time is not desirable. 
Common analytical approximations to the radiative transfer equation (RTE) fail to predict light transport, and could even give unphysical results. 
Therefore, we experimentally probe the position-dependent energy fluence rate of light inside scattering samples where the widely used $P_1$ and $P_3$ approximations to the RTE fail. 
The samples are aqueous suspensions of anisotropically scattering and both absorbing and non-absorbing spherical scatterers, namely microspheres ($r = 0.5~\rm{\mu}$m) with and without absorbing dye. 
To probe the energy fluence rate, we detect the emission of quantum dot reporter particles that are excited by the incident light and that are contained in a thin capillary. 
By scanning the capillary through the sample, we access the position dependence. 
We present a comprehensive discussion of experimental limitations and of both random and systematic errors. 
Our observations agree well with the Monte Carlo simulations and the $P_3$ approximation of the RTE with a correction for forward scattering. 
In contrast, the $P_1$ and the $P_3$ approximations deviate increasingly from our observations, ultimately even predicting unphysical negative energies. 
\end{abstract}

\maketitle
\section{Introduction} \label{sec:Probe3D_Intro}

\begin{figure}
    \centering
    \includegraphics[width=1.0\columnwidth, keepaspectratio]{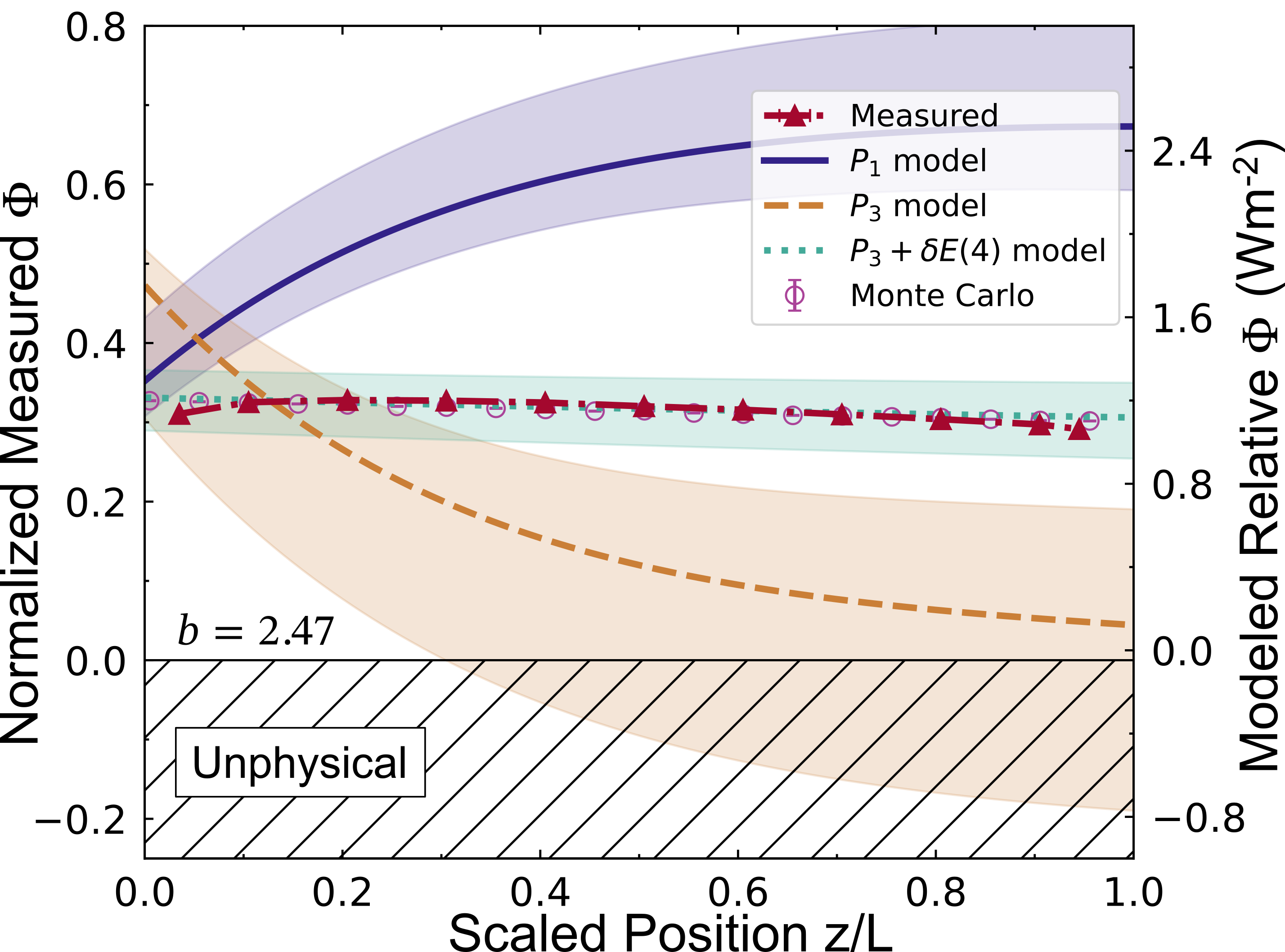}
    \caption{Measured (red triangles) and modeled energy fluence rate $\Phi (z)$ as a function of depth $z$, for a sample with optical thickness $b = \mu_{\rm{ext}}(\lambda) L = 2.47$.
    The measured $\Phi(z)$ is normalized to the incident intensity. 
    The curves represent the $P_N$ approximations to the radiative transfer equation, and the circles are Monte Carlo simulations. 
    The \textit{unphysical} range with \textit{negative} $\Phi$ is highlighted with black hatched lines.
    }
    \label{fig:DS3_meas_vs_models}
\end{figure}
For myriad applications such as light emitting diodes (LEDs)~\cite{Schubert2006Book, Krames2007JDT, Bechtel2008Conference, Akasaki2015RMP, Amano2015RMP, Nakamura2015RMP, Meretska2019ACSPhot, Lin2023JPhys}, atmospheric aerosols~\cite{Nousiainen2015Book}, oceanography and remote sensing~\cite{Funk1973AO, Vos1986CSEWM, Fournier1994OOXII, Fell2001JQSRT, Stramski2004PO, McCormick2006Book,Salama2015RSE}, and biophysics~\cite{Star1989BookChp, Jacques1990JQE, Cheong1990JQE, Kienle2006PRL}, as well as for fundamental physics like phase-conjugated scattering~\cite{Paasschens1997PRA}, random lasing~\cite{Pierrat2007PRA}, correlated and inhomogeneous scattering media~\cite{Wang2014PRA, Vynck2016PRA, Tommasi2020PRA}, and metamaterials~\cite{Arruda2018PRA}, it is crucial to understand the light transport. 
Hence, it is vital to predict the position-dependent energy fluence rate that is proportional to the energy density. 
While Monte Carlo simulations are a powerful tool to study the energy distribution in scattering media~\cite{Prahl1989BookChp, Mujumdar2010JN, Jacques2011BookChp, Atif2011OS, Uppu2013PRA, Leino2019OSAC, Jonsson2020OE, Cooper2021C, Krieger2021AA}, they come at the cost of extremely long computation times~\cite{Shonkwiler2009Book}. 
Analytical approximations to transport theory offer a much faster alternative~\cite{Joseph1976JAS, Star1988PMB, Keijzer1988AO, Dickey1998PMB, Egger2014AA, Liemert2011PRA, Liemert2021JOSAA, Ott2022JQSRT}, but they may not provide accurate results for samples that exhibit strong anisotropic scattering and absorption~\cite{Akdemir2022PRA}.

Experimental observations of the position-dependent energy fluence rate in \textit{real} scattering samples are essential to identify reliable methods to model the light transport in real devices. 
Such \textit{in situ} and \textit{in vitro} measurements of light transport have been widely made in different research areas, including aerosols~\cite{Dolgos2014OE, Boiger2022JAS}, LEDs~\cite{Chen2014IEEETDM}, and photodynamic therapy (PDT)~\cite{Star1987PP, Pomerleau2006PMB, Zhu2018BookChp}.

In Ref.~\cite{Akdemir2022PRA}, we have studied several analytic models of light transport that are summarized in Appendix~\ref{appx:TheoryBackground}, and presented their ranges of validity as well as their accuracies, as gauged by Monte Carlo simulations. 
As a next step, we decided to observe what happens \textit{in real samples} where analytic models break down. 
As an illustrative example, Figure~\ref{fig:DS3_meas_vs_models} shows the measured and modeled energy fluence rate $\Phi (z)$ of a forward scattering and moderately absorbing sample that lies in the unphysical range of the $P_3$ approximation to the radiative transfer equation.
Naively one would solely trust Monte Carlo simulations, but it is important to note that even the best simulation cannot fully replicate experimental observations. 
For instance, Monte Carlo simulations of light transport rely on input parameters such as scattering and absorption coefficients, which are typically interpreted from other experiments that carry additional assumptions and errors. 
Hence, accurate experimental observations are ultimately indispensable for device applications in, for instance, semiconductor metrology, solid-state lighting, or space observation optics, all of which are actual applications of our research program~\cite{FFSO2016}.

\begin{figure}[ht!]
    \centering
    \includegraphics[width=0.85\columnwidth, keepaspectratio]{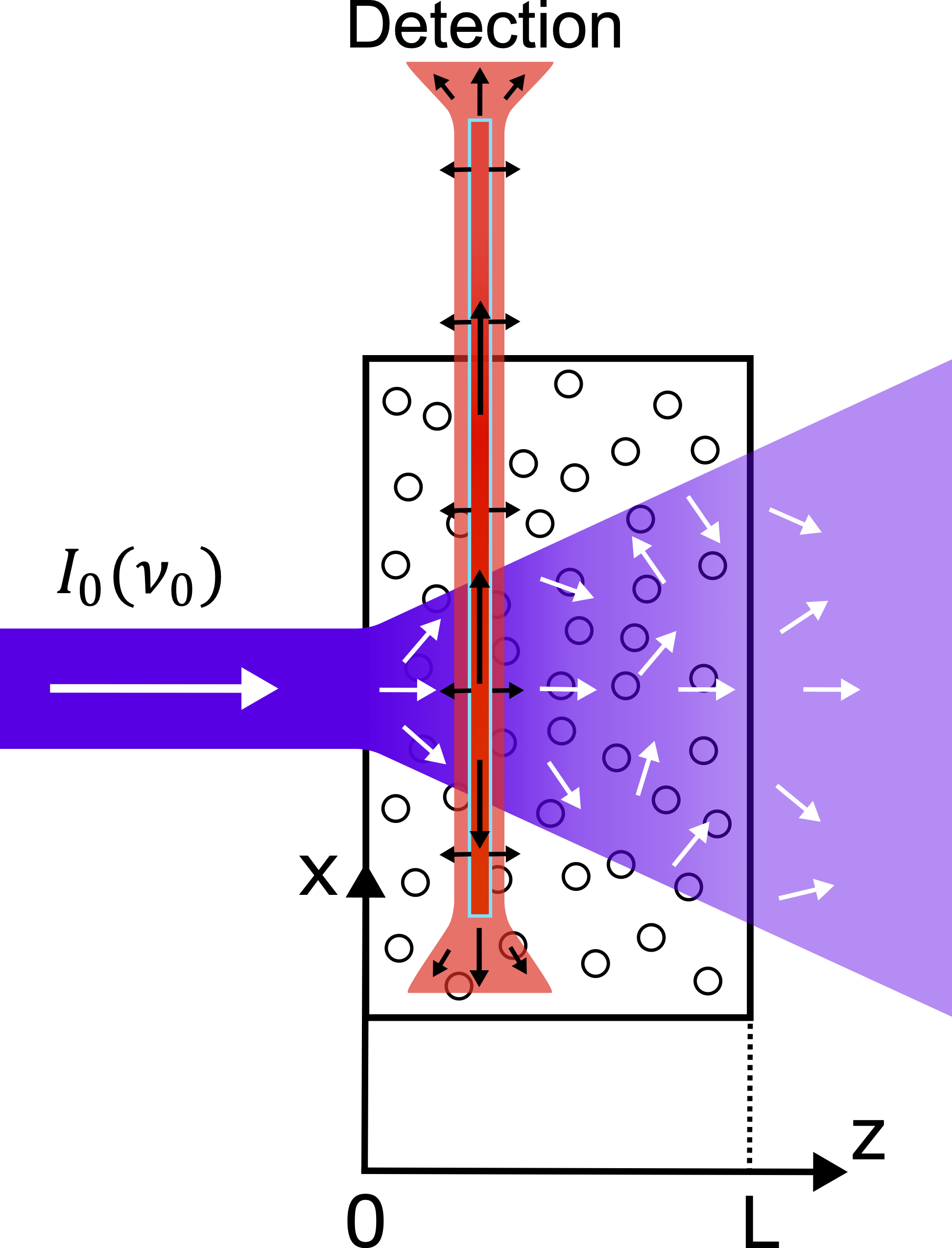}
    \caption{Cross-section of a slab with thickness L that contains an aqueous suspension of microspheres.
    A thin capillary filled with fluorescent reporters is inserted in the sample as a probe.
    The slab is illuminated from the side by light with intensity $I_0 (\nu_0)$.
    The white arrows represent the direction of light with frequency $\nu_{0}$, including both the diffuse and the unscattered parts.
    The black arrows represent the direction of emitted light with frequency $\nu_{\rm{e}}$.
    Fluorescent emission is detected from the top of the probe.
    }
    \label{fig:SuspensionSample_TopDetection}
\end{figure}

In this work, we report on experiments done in parameter ranges where most of the analytic approximations fail, to observe what happens and how well models pertain. 
The energy fluence rate $\Phi(\mathbf{r}, \nu)$ represents the total optical power through a unit spherical area, with units $\rm{Wm^{-2}Hz^{-1}}$. 
We simplify the representation of the 3D position $\mathbf{r}$ to a 1D $z$-dependence by considering the total rate versus $z$ position in a cross-section, see Fig.~\ref{fig:SuspensionSample_TopDetection}. 
For notational ease, we omit the frequency $\nu$ dependency to write the position-dependent energy fluence rate as $\Phi(z)$, since our study is a nearly monochromatic one. 
We present our experiments on the \textit{in situ} detection of the position-dependent fluence rate $\Phi (z)$ in both absorbing and non-absorbing, anisotropically scattering samples. 
We compare the experiments to well-known theoretical models, namely the $P_1$, $P_3$, and $P_3 + \delta E(4)$ approximations to the radiative transfer equation, and to Monte Carlo simulations of light transport, where the method of simulations used here is adapted from Ref.~\cite{Jacques2011BookChp}. 

A schematic of our experimental method is shown in Figure~\ref{fig:SuspensionSample_TopDetection}, where a slab with spherical scatterers is illuminated from the ($x, y$) side by light with intensity $I_0 (\nu_0)$.
Light enters the sample and propagates inside, undergoing scattering and absorption by the spheres. 
A probe consisting of a thin cylindrical capillary, filled with quantum dot reporter particles, is utilized to detect the intensity at known positions inside the sample, see, \textit{e.g.}, Refs.~\cite{Horstmeyer2015NatPhot, Hong2018Optica, Daniel2019OpEx}. 
When the light reaches the probe from any direction, it excites the quantum dot reporters that subsequently emit light at a Stokes-shifted lower frequency $\nu_{\rm{e}}$. 
Part of the re-emitted light is then collected from the top of the probe, and the position-dependent energy fluence rate is inferred from this measurement. 
Our observations match well with \textit{physical} and \textit{accurate} models for these types of samples, and show a clear distinction from the ones that fail and predict \textit{unphysical} negative energy fluence rates.

\section{Experimental details} \label{sec:ExperimentalDetails}
The experiments are done on samples with strongly forward-scattering (anisotropy $g > 0.75$) and absorbing (albedo $a < 1$) scatterers, to be in the parameter range where common analytical approximations fail~\cite{Akdemir2022PRA}.

\subsection{Sample preparation}\label{subsec:SamplePrep3D}
\begin{figure}
    \centering
    \includegraphics[width=0.9\columnwidth, keepaspectratio]{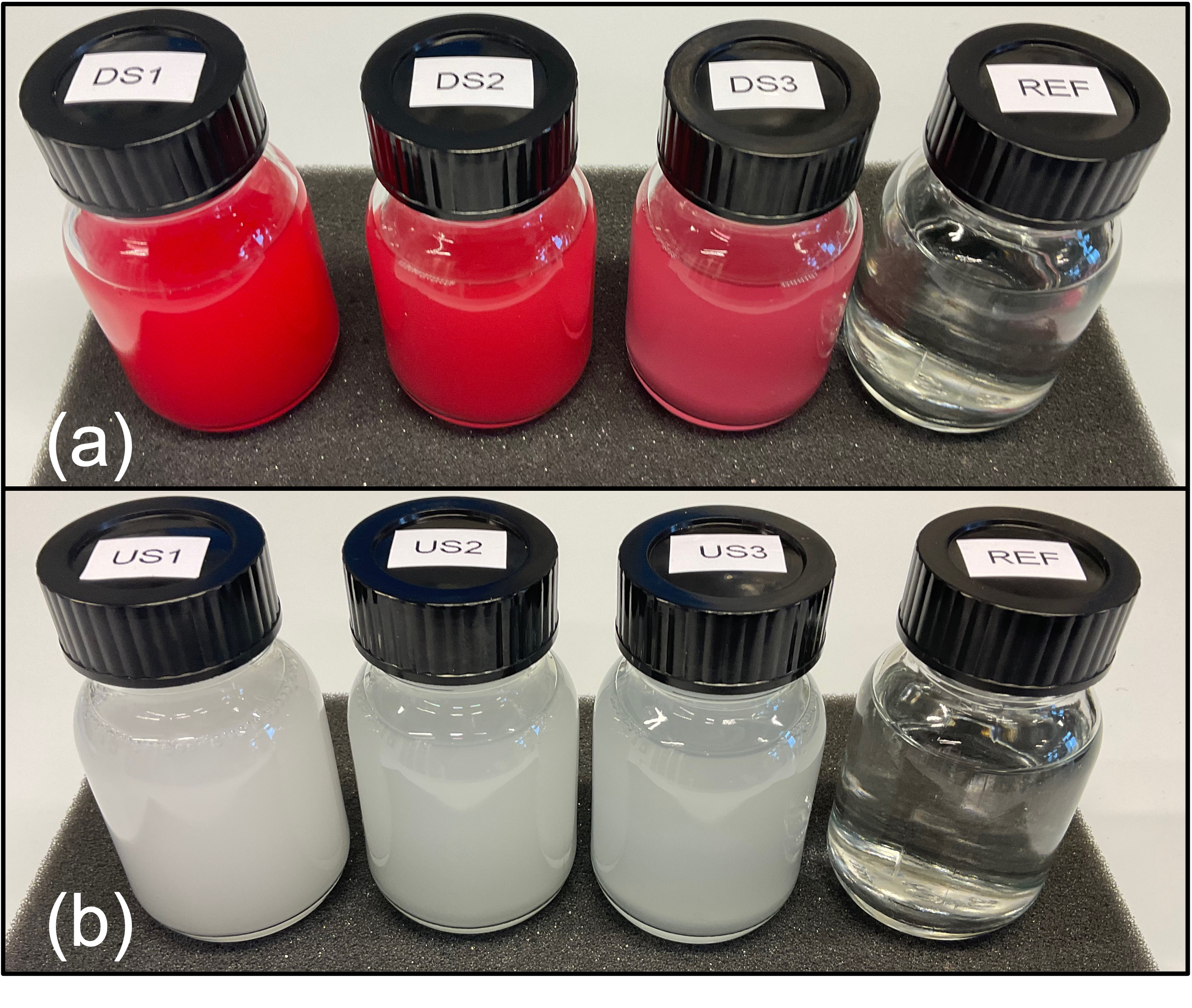}
    \caption{Photographs of the suspensions with different densities of (a) red dyed scatterers, (b) undyed scatterers.
    The scatterers are polystyrene spheres with $D = (1.00 \: \pm \: 0.05)~\rm{\mu m}$ diameter.
    }
    \label{fig:Samples}
\end{figure}
Samples are prepared by diluting commercially-bought polystyrene microsphere suspensions, with reported particle radii $r = 0.5 \pm 0.025$ $\rm{\mu m}$.
The microsphere suspensions used in our experiments are both plain and red-dyed Polybead\textsuperscript{\tiny\textregistered} Polystyrene Microspheres from Polysciences for undyed and dyed sphere suspensions, respectively.
A buffer solution is produced by mixing Sodium dodecyl sulfate (SDS) surfactant ($ 0.05\% $) with de-ionized (DI) water ($ 99.95 \% $).
The samples with scatterers are then prepared by diluting their respective suspensions with the buffer solution, 
where syringe filters with 0.2 $\rm{\mu m}$ pore size are used to remove residual particles from the buffer solution. 
The use of SDS in the buffer solution serves to stabilize the colloidal suspension, and prevent agglomeration and sedimentation of the scatterers. 
The samples prepared with this buffer solution are stable for at least 24 hours, after which some initial sedimentation is observed. 
Prior to experiments, the samples are carefully agitated, with attention paid to prevent the introduction of air bubbles.
Overall, and based on our previous detailed colloidal studies~\cite{Megens1997Langmuir, Riese2000PRL, Megens2001PRL, Hartsuiker2008Lang}, we estimate the fraction of non-single spheres to be less than about $5 \%$, hence Mie theory for single spheres is valid for our samples. 
Dyed sphere samples are labelled DS, and plain (undyed) sphere samples are US. 
A photograph of the samples is shown in Figure~\ref{fig:Samples}, showing the different diffuse color appearance of the DS samples and different whiteness of the US samples. 
The contents and scatterer concentrations of prepared samples are listed in Table~\ref{tab:SampleProperties}. 
Since the weight fraction of polystyrene spheres in water is closely equal to the volume fraction, and since all our suspensions have a volume fraction less than $0.026~\mathrm{vol}\%$, the structure factors of our suspensions are very nearly equal to $S(Q) = 1$~\cite{Vos1991book}, hence Monte Carlo simulations of light transport do not require structural corrections~\cite{Rojas-OchoaPRL2004}. 

\begin{table*}[ht] 
    \begin{adjustbox}{width=0.6\textwidth}
        \begin{tabular}{lcccc}
    			\hline
                \hline
                \textbf{Sample name} & & \textbf{Contents} & &  $\boldsymbol{\rho} \: \mathbf{(\rm{mm}^{-3})}$\\
                \hline
                Reference & & $99.95 \: wt.\%$ DI water + $0.05 \: wt.\%$ SDS & & 0 \\
    			\hline
                DS1 & & $0.025\% \: (w/v)$  dyed sphere + Ref & & $(4.6 \: \pm \: 0.6) \times 10^5$ \\
                \hline
                DS2 & & $0.013\% \: (w/v)$ dyed sphere + Ref & & $(2.3 \: \pm \: 0.3) \times 10^5$ \\
                \hline
                DS3 & & $0.006\% \: (w/v)$ dyed sphere + Ref & & $(1.1 \: \pm \: 0.1) \times 10^5$ \\
                \hline
                US1 & & $0.026\% \: (w/v)$ plain sphere + Ref & & $(4.7 \: \pm \: 0.6) \times 10^5$ \\
                \hline
                US2 & & $0.013\% \: (w/v)$ plain sphere + Ref & & $(2.4 \: \pm \: 0.3) \times 10^5$ \\
                \hline
                US3 & & $0.007\% \: (w/v)$ plain sphere + Ref & & $(1.2 \: \pm \: 0.2) \times 10^5$ \\
                \hline
                \hline
    	\end{tabular}
    \end{adjustbox}
\caption{Contents and the number density of scatterers $\rho$ of the prepared samples.
The percentages of spheres are adjusted from the information provided by the vendor about their products. 
DS are red-dyed microsphere samples, and US are undyed microsphere samples.
}
\label{tab:SampleProperties}
\end{table*}

\subsection{Experimental Setup} \label{subsec:ExperimentalSetup_3D}
\begin{figure}[h!] 
    \centering
    \includegraphics[width=1.0\columnwidth, keepaspectratio]{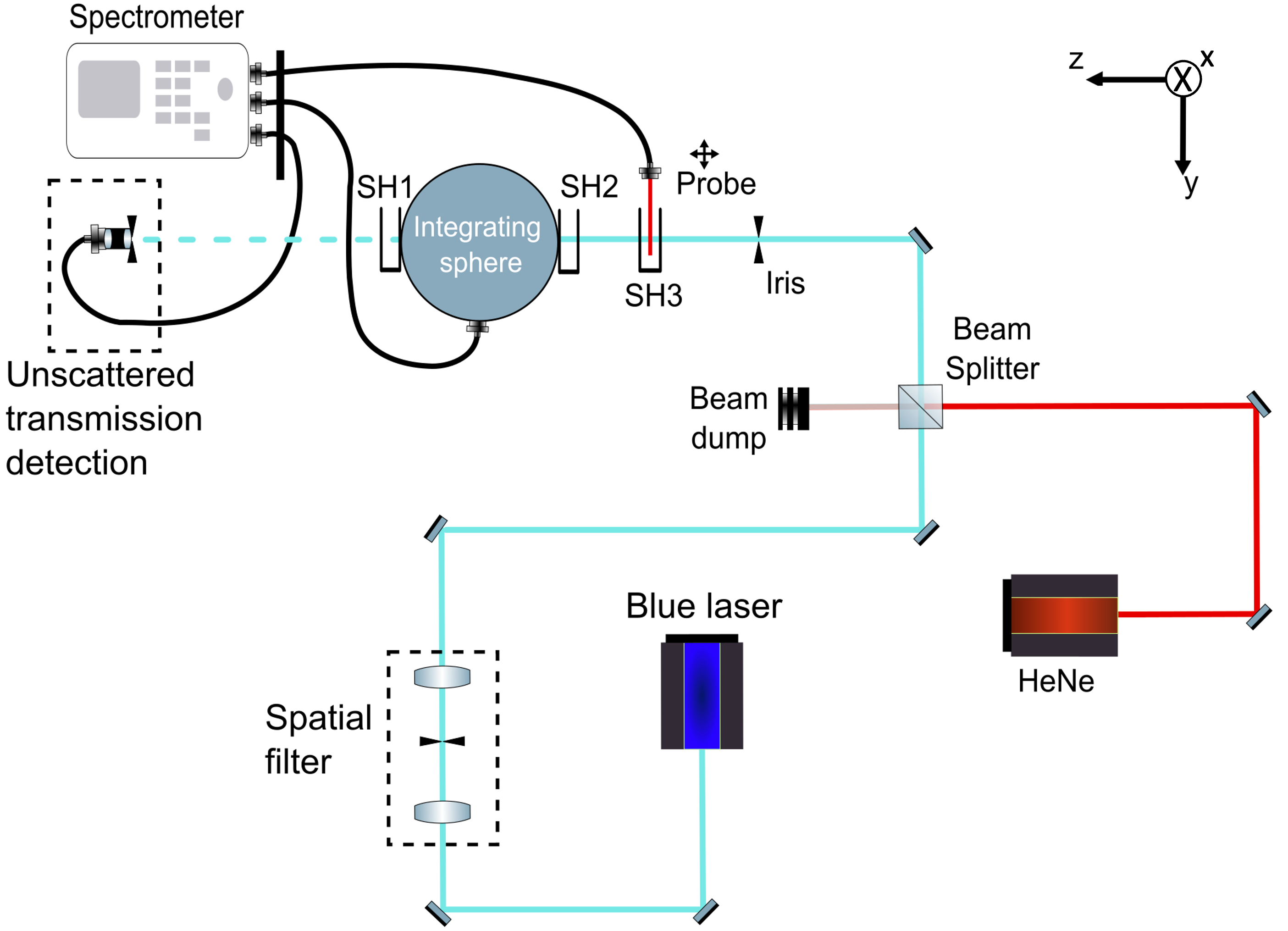}
    \caption{Schematic of the experimental setup: the blue laser ($\lambda = 408.5$ nm) is the primary light source that is spatially filtered and combined with the HeNe laser at the beam splitter for alignment purposes. 
    The probe is attached to an assembly of motorized translation stages for precise positioning in 3 dimensions.
    The spectrometer is connected to the probe, integrating sphere, and unscattered transmission detection via fiber-optic cables.
    Several components, including ND filters and beam dumps, have been omitted for clarity. 
    }
    \label{fig:SetupSketch}
\end{figure}

Figure~\ref{fig:SetupSketch} presents a schematic of the setup used in our experiments.
A blue laser (Thorlabs NPL41B, $\lambda = 408.5$ nm) is used as the primary light source, see Appendix~\ref{subsec:ChoiceLightSource}. 
The laser beam is spatially filtered because multiple modes inside the laser cavity are not collimated as well as the fundamental mode, while they perturb the beam profile at the sample. 
A substantial improvement is observed upon spatial filtering, see Ref.\cite{Akdemir2023Thesis}. 

%

\begin{figure}
    \centering
    \includegraphics[width=1.0\columnwidth, keepaspectratio]{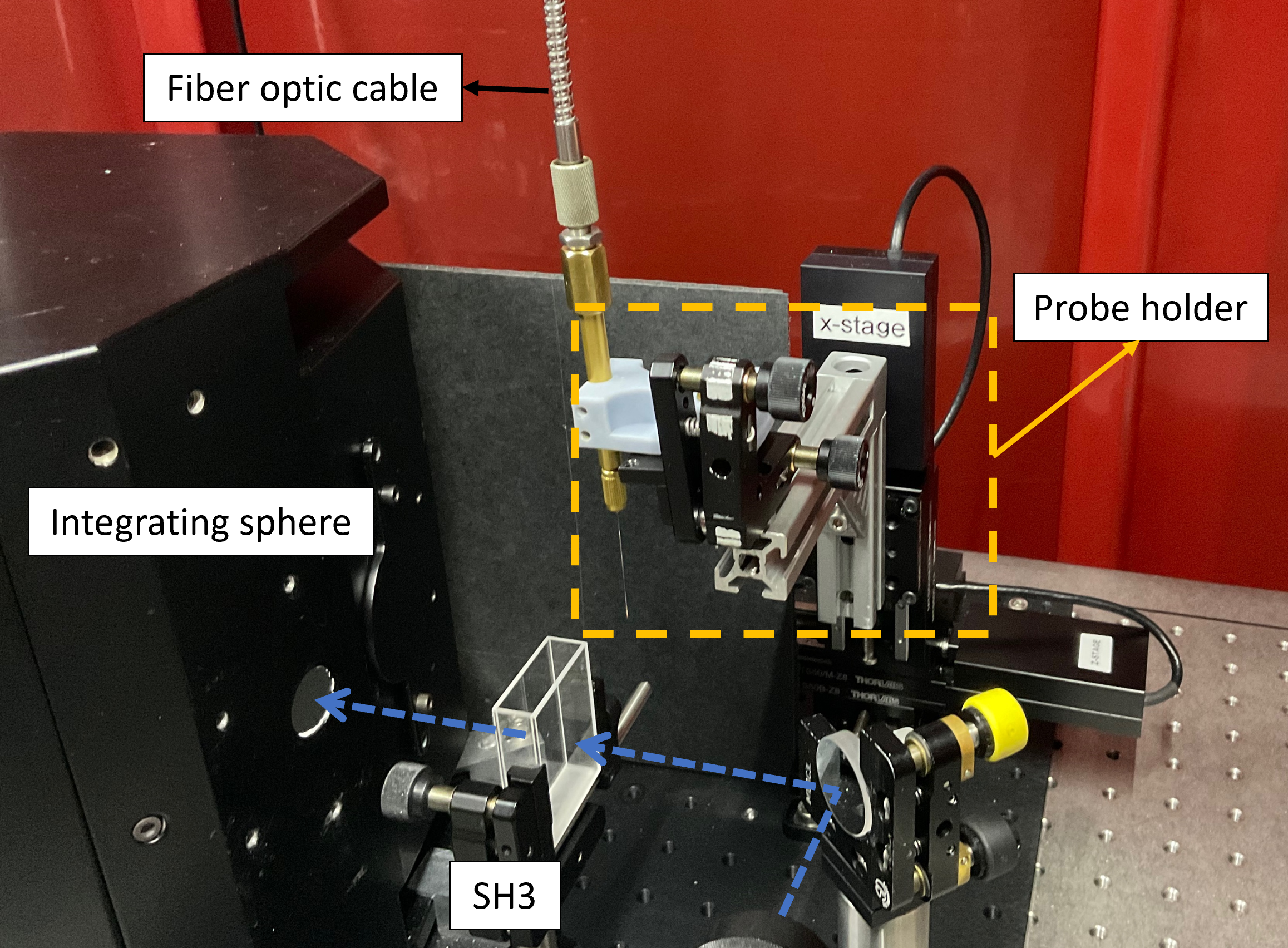}
    \caption{Photograph of the assembled probe holder.
    The fiber attached to the probe, the integrating sphere transmission port and the sample holder SH3 with an empty cuvette attached to it, is also visible in the photo.
    }
    \label{fig:ProbeHolder}
\end{figure}

The sample holders SH1 and SH2 are attached to the integrating sphere, whereas the holder SH3, where the probe is immersed in (see Fig.~\ref{fig:SetupSketch} and Fig.~\ref{fig:ProbeHolder}), is positioned separately from the integrating sphere. 
The probe is connected to a custom-built holder with 3D-printed and manually made parts, as shown in Figure~\ref{fig:ProbeHolder}. 
For details on the preparation of the probe, see appendix~\ref{subsec:Probe_3D}. 
To detect the intensity emitted from the probe, a connector part is created in the workshop to connect the probe with a fiber-optic cable that is attached to a spectrometer (Avantes Starline, AvaSpec-2048L) on the other end.
The holder is attached to an assembly of motorized translation stages (Thorlabs MTS50/M-Z8) that moves the probe to precise positions in 3 dimensions.
In-house Python scripts are used to control the movement of the probe and detection with the spectrometer.
The $x$ direction movement is utilized to move the probe in and take it out of the sample, whereas the y and $z$ direction movements are used to scan the sample and measure the intensity at specific positions.
Total transmission and total reflection measurements are done with an integrating sphere (Opsira UKU240), with a port diameter $d_{\rm{port}} = 20$ $\rm{mm}$ for both transmission and reflection ports.

Before collecting data, each sample suspension is transferred to a wide quartz cuvette (outer dimensions $12.5 \times 32.5 \times 45 \: \rm{mm}$) with an inner thickness of $L = 10$ $\rm{mm}$ along the $z$ direction, as is shown in Figure~\ref{fig:ProbeHolder}. 
These cuvettes are then placed in sample holders for measurements.
The holders SH1 and SH2 are used for total reflection and total transmission measurements, respectively.
Holder SH3 shown in Fig.~\ref{fig:ProbeHolder} is used in the measurements with the probe and to measure unscattered transmission ($T_{\rm{u}}$) that represents the light that survives extinction while maintaining its original direction.  
The distance between the fiber coupler of $T_{\rm{u}}$ detection and SH3 is 750 mm, which results in a half-acceptance angle $\theta = 0.04^{\circ}$.
The fraction of detected \textit{scattered} light is estimated to be $< 0.1 \%$ in our $T_{\rm{u}}$ measurements~\cite{Li2016AO}.

\section{Results and discussion} \label{sec:ResultsAndDiscussion_Probe3D}

In this section, the results of unscattered transmission, total transmission, total reflection, and probe measurements are discussed.
In addition, the method to extract the albedo $a$ and the anisotropy $g$ of the samples is discussed.

\subsection{Unscattered Transmission, Total Transmission and Total Reflection} \label{subsec:Tu_TT_TR_3D}

The unscattered transmission $T_{\rm{u}}$ represents the portion of the incident light that is transmitted through the sample without deviating from its initial direction of travel. 
Usually, the Beer-Lambert-Bouguer's law is used to express $T_{\rm{u}}$ as
\begin{equation}
\label{eqn:Tu_BeerLambert}
	T_{\rm{u}} (\lambda, \rho) = \frac{I_{\rm{u}}(\lambda, \rho)}{I_0(\lambda)} = e^{-\rho \sigma_{\rm{ext}}(\lambda, \rho) L}.
\end{equation}
Here, $I_{\rm{u}}$ is the intensity transmitted without changing its direction, $I_0$ is the incident intensity, $\rho$ the number density, $\sigma_{\rm{ext}}$ the extinction cross-section, and $L$ the sample thickness.
Furthermore, the optical thickness $b$ is defined to be 
\begin{equation}
\label{eqn:optThickness}
	b(\lambda, \rho) \equiv \rho \sigma_{\rm{ext}}(\lambda) L = \mu_{\rm{ext}}(\lambda) L,
\end{equation}
where $\mu_{\rm{ext}}$ is the extinction coefficient.
\begin{figure}[h!]
    \centering
    \includegraphics[width=1.0\columnwidth, keepaspectratio]{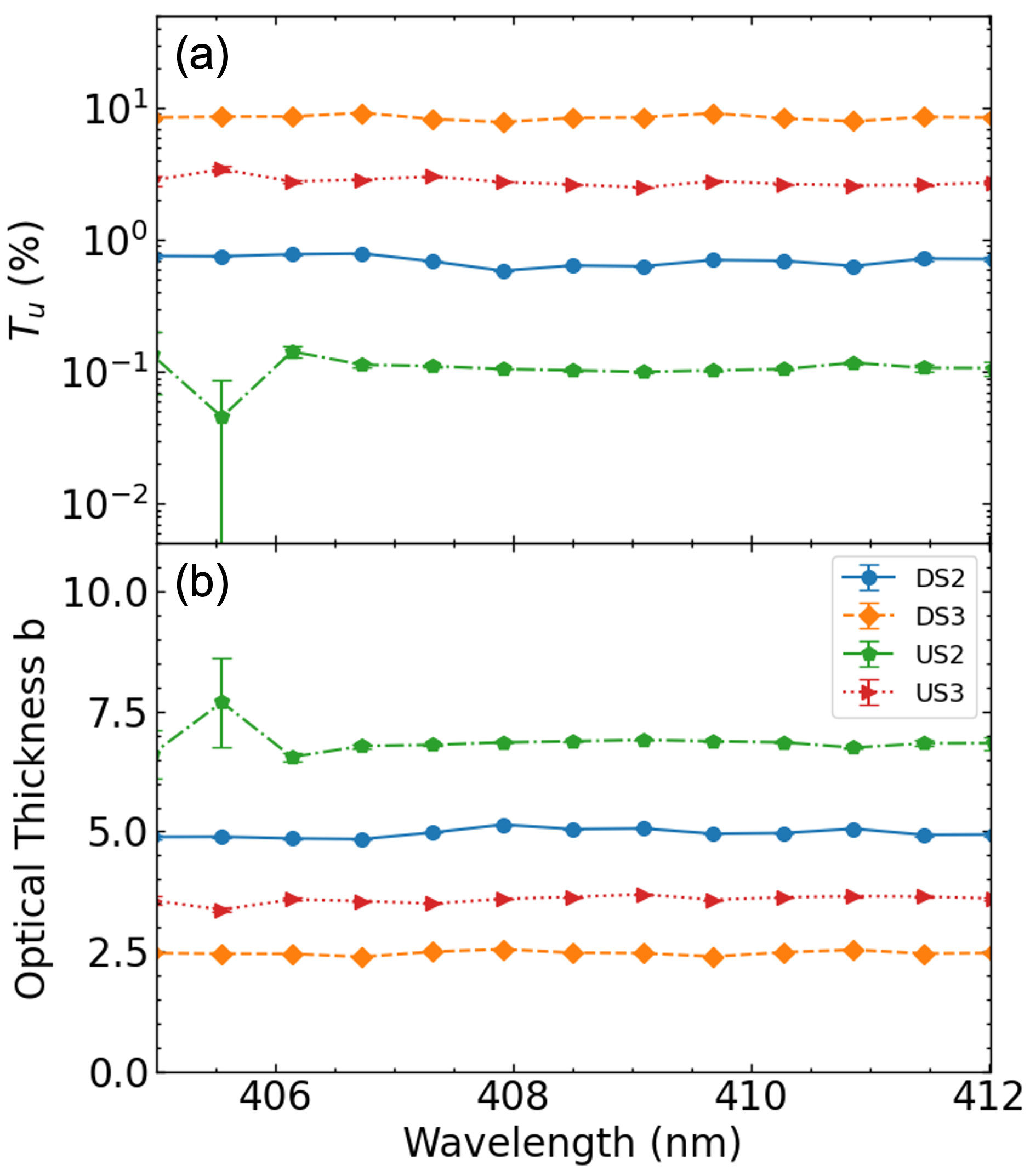}
    \caption{(a) Unscattered transmission $T_{\rm{u}}$, and (b) the extracted optical thickness $b$ of samples DS2, DS3, US2 and US3.
    }
	\label{fig:Tu_optT}
\end{figure}
Figure~\ref{fig:Tu_optT}(a) presents the measured unscattered transmission $T_{\rm{u}}$, and Figure~\ref{fig:Tu_optT}(b) displays the extracted optical thickness (see Eq.~\ref{eqn:optThickness}) of samples DS2, DS3, US2, and US3 throughout the bandwidth of the blue laser.
The densest samples DS1 and US1 have a vanishing transmission below our detection limit and are not shown. 
Therefore, the optical thickness of DS1 and US1 are calculated from the measurements of other samples, by appropriately scaling the density of scatterers. 
The results show a clear effect of the density of scatterers and the composition of the scatterers on the light transmission.
Naively, one would expect dyed samples to be optically thicker than undyed samples with similar scatterer concentrations due to increased absorption, but our measurements show the opposite behavior. 
We attribute this intriguing result to the presence of the unknown red dye inside the scatterers that alters the refractive index of the polystyrene spheres in DS2 and DS3. 
The dye likely increases the anisotropy, causing the spheres to scatter light more in the forward direction.\footnote{We performed Mie scattering calculations for relevant parameters, namely sphere with diameter $D = 1.0~\rm{\mu m}$ and real part of the refractive index $n_{\rm{s}}' = 1.61$, in a medium with $n_{\rm{out}} = 1.33$ typical of water at wavelength $\lambda = 0.4085~\rm{\mu m}$, for increasing imaginary sphere's index $n_{\rm{s}}'' = 0.0004$ (undyed), $0.01$, $0.1$, $0.2$, resulting in $g = 0.919$, $0.924$, $0.950$, $0.954$, respectively, thus agreeing with our hypothesis. 
If we further increase the imaginary part to $n_{\rm{s}}'' = 1$ (unrealistic) the anisotropy decreases to $g = 0.871$. 
For reference, in case of gain $(n_{\rm{s}}'' < 0)$~\cite{Olmos-Trigo2020PRL}, we find that $g$ decreases increasingly rapidly, to a point where $g < 0$, the albedo becomes absurdly large, and a divergence in Mie resonance linewidth occurs~\cite{vanderMolen2006OL}.}
Consequently, the unscattered part of the transmitted light increases, which explains our observations in Figure~\ref{fig:Tu_optT}. 
%
\begin{figure}
    \centering
    \includegraphics[width=1.0\columnwidth, keepaspectratio]{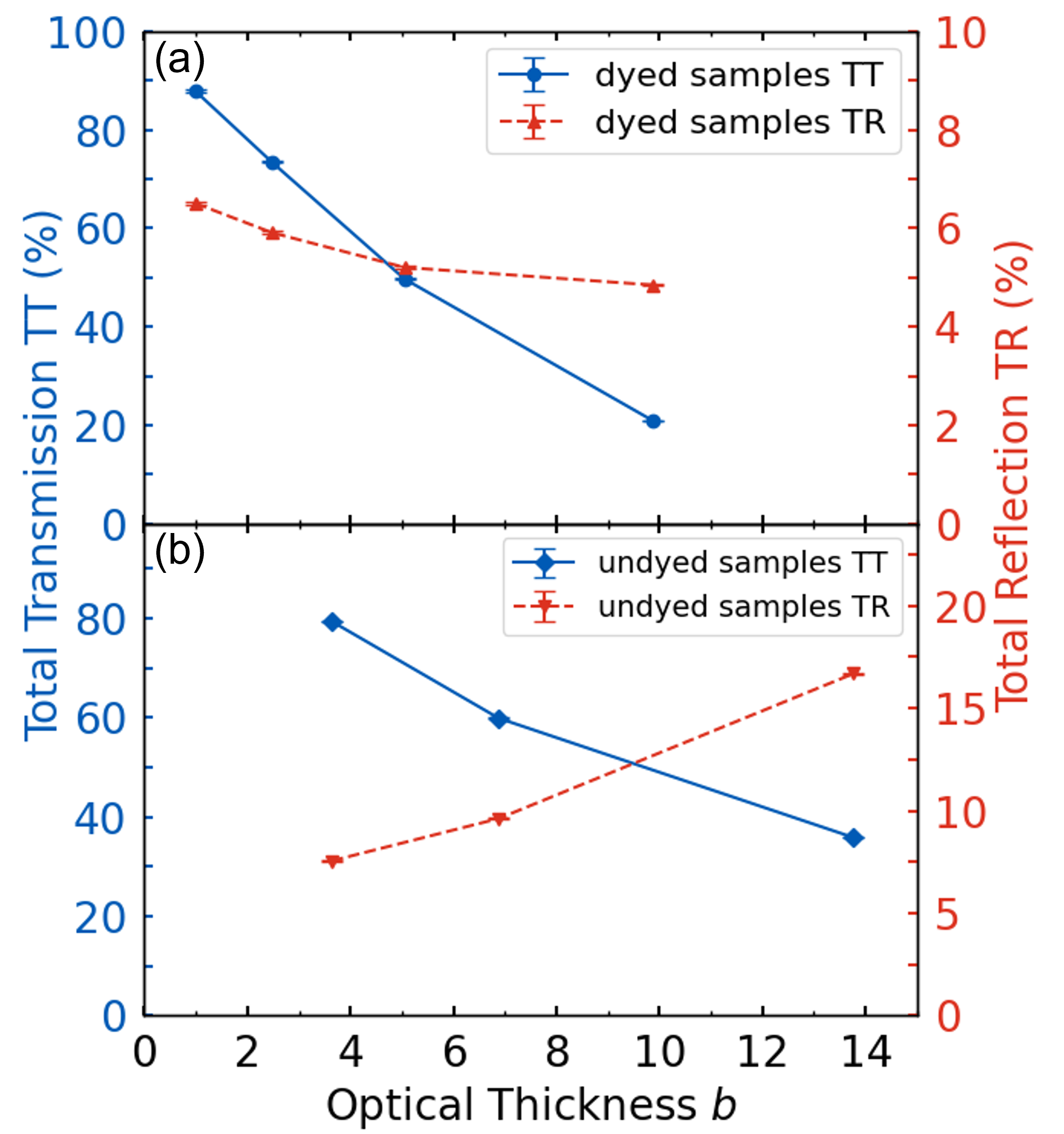}
    \caption{Total transmission TT($b$) and total reflection TR($b$) as a function of optical thickness for (a)dyed microsphere suspensions, (b) and for undyed microsphere suspensions. 
    The lines connect the data points. 
    }
    \label{fig:TT_TR_3D}
\end{figure}

Total transmission (TT) and total reflection (TR) of dyed and undyed microsphere suspensions are displayed in Figure~\ref{fig:TT_TR_3D} as a function of their optical thicknesses at the central wavelength $\lambda = 408.5$ nm. 
As the optical thickness $b$ increases, both TT and TR decrease for the dyed microspheres, see Figure~\ref{fig:TT_TR_3D}(a). 
The decrease in TT matches with increased absorption and increased scattering. 
The simultaneous decrease in TR agrees with the increasing density of forward scattering microspheres and the increasing absorption when the suspensions get optically thicker. 

Since the scatterers in the dyed microsphere suspensions are filled with an unknown red dye (the manufacturer was unable to share the information with us), we cannot use Mie theory to obtain the albedo and anisotropy $(a, g)$, since the complex refractive index of the scatterers is unknown. 
Therefore, our TT and TR measurements are necessary to determine the $(a, g)$ scattering parameters of the dyed samples using models such as $P_{\rm{N}}$ approximations to the RTE, and Monte Carlo simulations of light transport. 
While the random errors in the measurements are negligibly small, see Figure~\ref{fig:TT_TR_3D}, systematic errors merit attention during the determination of the scattering parameters. 

For the samples with dyed scatterers, systematic errors are more significant at low scatterer density, where the absorption is low. 
The reason is that the suspensions are held in quartz cuvettes, and escaping light reflects multiple times from the cuvette walls before leaving the sample. 
Hence, light exits the sample and cuvette at wider angles, corresponding to a increased diffuse spot whose size exceeds the port sizes of the integrating sphere. 
This effect is notable in total reflection, since here light is incident on the sample at $6^\circ$ off the normal due to the (generic) design of the integrating sphere. 
If the sample absorption is strong, this effect becomes minor since the internally reflected light is quickly attenuated, hence the angular spread is not substantial. 
Since errors are largest at lowest densities, and since we know that the zero-density reference sample (buffer solution with SDS) has $a = 1$ and $g = 1$, we estimate maximum systematic errors in both TT and TR by comparing the measurements to the Monte Carlo simulations for this reference sample. 
This comparison yields maximum systematic errors $\Delta TT = 6\%$ in total transmission and $\Delta TR = 4\%$ in total reflection. 
We utilize these errors to extract a range of possible $(a,g)$ parameters for each dyed sample, and we take the average of that range as the extracted $(a,g)$ pair. 
While the determination of the $(a,g)$ parameters is best for the densest dyed samples, we will see below that the parameters obtained from all samples are in good mutual agreement. 

For undyed sphere suspensions, Figure~\ref{fig:TT_TR_3D}(a) shows that TT decreases with increasing scatterer density (or increasing optical thickness), as expected, see \textit{e.g.}~\cite{vanRossum1999RMP, Akkermans2007Book, Vos2013AO}. 
In contrast to the dyed microspheres, Figure~\ref{fig:TT_TR_3D}(b) shows that TR increases with increasing optical thickness, since the light that is less transmitted (compare Fig.~\ref{fig:TT_TR_3D}(a)) is necessarily diffusely reflected, and since absorption is negligible, no light gets lost in the bulk of the samples. 

From the reasoning above about reflected light escaping undetected in weakly absorbing samples, we conclude that the systematic error in TR is greater for undyed sphere samples than for the dyed spheres, since the amount of light escaping at even greater angles has increased further. 
As an illustration, the theoretically computed total reflection for sample US2 is $TR_{\mathrm{P_3 + \delta E(4)}} = 28.4 \%$, whereas the measured total reflection is much less namely $TR_{\mathrm{exp}} = 10 \%$. 
In contrast, the systematic error in total transmission is much less, as illustrated by theoretical total transmission for US2 being $TT_{\mathrm{P_3 + \delta E(4)}} = 65.6 \%$, which is in good agreement with the measured total transmission $TR_{\mathrm{exp}} = 61 \%$. 
Therefore, even though TT agrees well, our extraction method is not able to obtain $(a,g)$ parameters for undyed spheres, since TR is substantially off. 
Therefore, we employ Mie theory for undyed spheres. 

\subsection{Determining $(a, g)$ parameters of the samples} \label{subsec:Finding_a_g_pair}
In this section, the albedo and anisotropy $(a, g)$ of dyed microsphere suspensions are obtained by comparing TT and TR measurements to Monte Carlo simulations of light transport, using a \textit{brute-force} approach. 
We emphasize that internal reflections from cuvette walls are included in the Monte Carlo simulations to obtain results as close as possible to our measurements.
First, the optical thicknesses $b$ of the samples are extracted, and then Monte Carlo simulations of light transport are performed for a grid of $(a, g)$ pairs covering all possible albedos and anisotropies.
We then use the relative cost function $S(a,g)$ defined as
\begin{equation}
S(a,g) \equiv \sqrt{\delta TT (a, g, a_0, g_0) + \delta TR (a, g, a_0, g_0)},
\end{equation}
where the squared relative distances $\delta TR (a, g, a_0, g_0)$ and $\delta TR (a, g, a_0, g_0)$ are equal to
\begin{equation}
\begin{split}
    \delta TT (a, g, a_0, g_0) &= \frac{(TT_{\rm{MC}}(a,g) - TT_{\rm{DS\#}}(a_0,g_0))^2}{TT_{\rm{DS\#}}^2 (a_0,g_0)}, \\
    \delta TR (a, g, a_0, g_0) &= \frac{(TR_{\rm{MC}}(a,g) - TR_{\rm{DS\#}}(a_0,g_0))^2}{TR_{\rm{DS\#}}^2 (a_0,g_0)}, 
\end{split}
\end{equation}
and where $TT_{\rm{MC}}(a,g)$ and $TR_{\rm{MC}}(a,g)$ are the total transmission and reflection from Monte Carlo simulations, $TT_{\rm{DS\#}}(a_0,g_0)$ and $TR_{\rm{DS\#}}(a_0,g_0)$ are measured, $(a,g)$ is the running coordinate in the albedo-anisotropy grid, and $(a_0,g_0)$ is the albedo-anisotropy pair of the sample that we want to extract.
For the measured $TT_{\rm{DS\#}}(a_0,g_0)$ and $TR_{\rm{DS\#}}(a_0,g_0)$ quantities, $\#$ represents the number of the sample name, given in Table~\ref{tab:SampleProperties}.

\begin{figure}
	\centering
	\includegraphics[width=1.0\columnwidth, keepaspectratio]{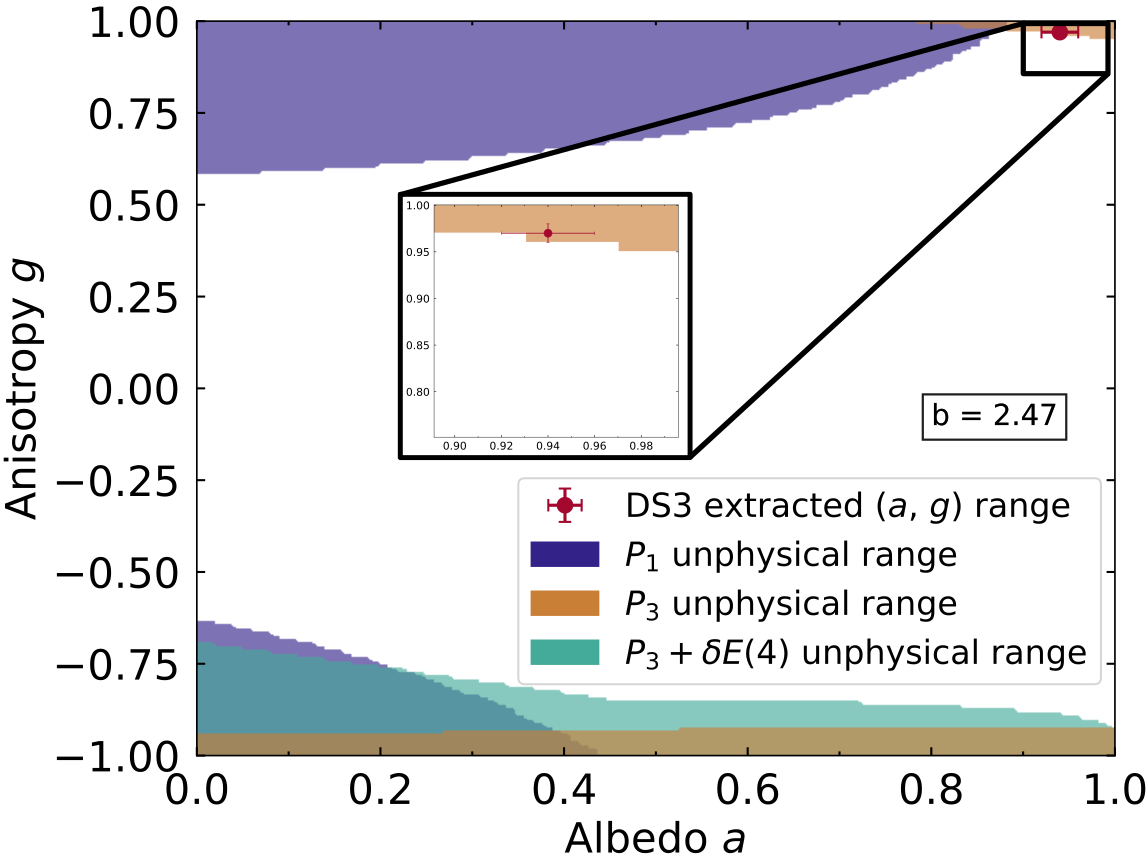}
	\caption{The extracted parameter range $(a,g)$ using the cost function $S(a,g)$ for sample DS3 with optical thickness $b = 2.47$, shown with the unphysical ranges of $P_1$, $P_3$, and $P_3 + \delta E(4)$ approximations. 
    The inset highlights the extracted $(a,g)$ typical of a forward scattering and absorbing sample. 
    }
    \label{fig:extracted_param_with_unphys_ranges}
\end{figure}

Figure~\ref{fig:extracted_param_with_unphys_ranges} shows the extracted parameters for sample DS3 ($b=2.47$) using the \textit{brute-force} method explained above.
In addition, Figure~\ref{fig:extracted_param_with_unphys_ranges} shows the unphysical ranges of $P_1$, $P_3$, and $P_3 + \delta E(4)$ approximations for the sample with optical thickness $b=2.47$. 
The horizontal and vertical axes represent all possible albedo $a$ and anisotropy $g$ values, respectively.
The minimum cost $S(a , g)$ (the best fit) is found, as expected, in the strong forward scattering and moderately absorbing region of the $(a,g)$ grid, which is in the \textit{unphysical range} of $P_3$ approximation~\footnote{Remarkably, \textit{physical} ranges do not imply better accuracy~\cite{Akdemir2022PRA}.}.
However, it is noteworthy that the best fit to our measurements is at $g_0=1$, which is an extreme point and is realistically unattainable in real samples with scatterers~\footnote{As the extreme points are not physical, they are not considered in $S(a,g)$ calculations.}.
This implies that errors in our measurements have a significant effect on the determination of the $(a,g)$ parameters of the samples.
We utilize the errors $\Delta TT$ and $\Delta TR$ reported in section~\ref{subsec:Tu_TT_TR_3D} to extract the $(a,g)$ parameters of the dyed samples.
The extracted $(a,g)$ parameters of the dyed microsphere suspension samples are given in Table~\ref{tab:Extracted_a_g}, which shows that the dyed samples are strongly forward scattering, and have a moderate absorption. 
We note that the mutual agreement between the $(a,g)$ parameters obtained for the different samples is very good, within $\delta (a) < 2\%$ for the albedo and $\delta (g)< 1\%$ for the anisotropy, which is even better than expected from the systematic errors discussed above. 

\begin{table}
\centering
    \begin{tabular}{lcccc}
			\hline
            \hline
            \textbf{Sample name} & & \textbf{Extracted $\mathbf{a}$} & &  \textbf{Extracted $\mathbf{g}$}\\
            \hline
            DS1 & & $0.90 \pm 0.02$ & & $0.95 \pm 0.02$\\
            \hline
            DS2 & & $0.91 \pm 0.01$ & & $0.96 \pm 0.02$ \\
            \hline
            DS3 & & $0.94 \pm 0.02$ & & $0.97 \pm 0.01$ \\
            \hline
            \hline
	\end{tabular}
\caption{Extracted albedo $a$ and anisotropy $g$ values of dyed microsphere suspension samples DS1, DS2 and DS3.
The values reported here are averaged over the extracted range of $(a,g)$ pairs, and errors are the standard deviation of the mean.
}
\label{tab:Extracted_a_g}
\end{table}
The comparison between simulated and measured transmission and reflection signals is not necessary for the undyed microsphere suspensions, as their optical properties are obtained directly from Mie theory for spheres. 
From previous characterization studies \cite{Megens1997Langmuir, Rates2023Thesis}, it is known that such microspheres are highly spherical, hence Mie theory that pertains to spheres gives an excellent description of the scattering properties.
The complex refractive index of polystyrene spheres at $408.5$ nm is taken from reference~\cite{Ma2003PhysMedBiol} to be $n_{\rm{poly}} = 1.61 + 0.0004i$, which yields non-absorbing $a = 0.997$ and strong forward scattering $g = 0.919$ for the undyed microspheres suspended in DI water. 
Mie calculations also provide the extinction cross-section $\sigma_{\rm{ext}}$, which we utilized together with the uncertainty of the scatterer radius $\Delta r_{\rm{scat}} = \pm 0.03$ $\rm{\mu m}$ to determine the $13\%$ uncertainty in the reported densities listed in Table~\ref{tab:SampleProperties}. 
The extracted $(a,g)$ parameters for both dyed and undyed samples lie within the range where both $P_1$ and $P_3$ approximations fail, whereas the $P_3 + \rm{\delta E}(4)$ approximation is expected to give accurate results. 

\subsection{Position-dependent energy fluence rate of dyed microsphere suspensions} \label{subsec:PositionDependentEnergy_3D_DyedSamples}
To determine the position-dependent energy fluence rate $\Phi(z)$, the light intensity at specific points inside samples is measured using the experimental setup described in subsection~\ref{subsec:ExperimentalSetup_3D}.
The coordinate system adopted in this section is depicted in Figure~\ref{fig:SetupSketch}.
The objective is to infer $\Phi(z)$ from measurements of the emission of quantum dots inside the probe, which are excited by the blue light inside the sample, along $z$ from the left boundary at $z = 0$ where the incident light enters, to the right boundary at $z = 10$ mm.
For the details of the probing procedure, see appendix~\ref{appx:ProbeProcedure}. 
The experimental data are compared to Monte Carlo simulations and analytical $P_1$, $P_3$, and $P_3 + \delta E(4)$ approximations to the radiative transfer equation~\cite{Akdemir2022PRA}.

%
\begin{figure}[b]
    \centering
    \includegraphics[width=1.0\columnwidth, keepaspectratio]{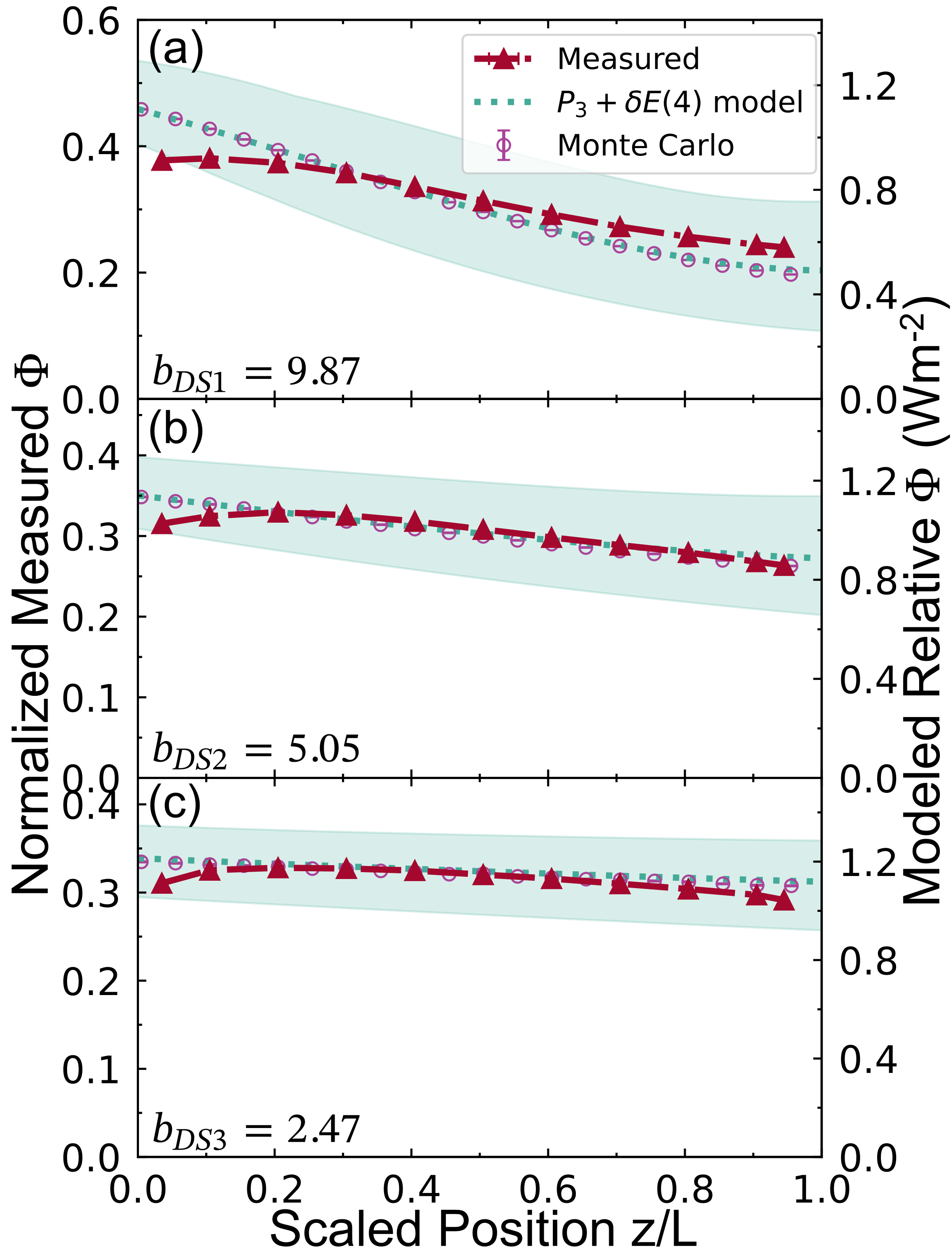}
    \caption{Comparison of measurement and models for samples (a) DS1, (b) DS2, and (c) DS3.
    The horizontal axis represents the $z$ positions scaled by the thickness of the sample.
    The vertical axis on the left displays the $\Phi(z)$ measurements of samples normalized to the measurements of the incident intensity.
    The right vertical axis corresponds to the modeled relative $\Phi(z)$.
    The dotted lines represent the calculated $\Phi(z)$ from the $P_{3}+\delta E(4)$ approximation.
    The dynamic range of the $P_{3} + \delta E(4)$ approximation is shown as the green area.
    The circles represent the results of Monte Carlo simulations, and the triangles connected with red dash-dotted lines represent the measured results of samples.
    The measured results are calculated by integrating the $y$ scans at each $z$ position, shown as triangles.
    }
    \label{fig:ModelVsMeasurement_DS_samples}
\end{figure}
\begin{figure}[b]
    \centering
    \includegraphics[width=1.0\columnwidth, keepaspectratio]{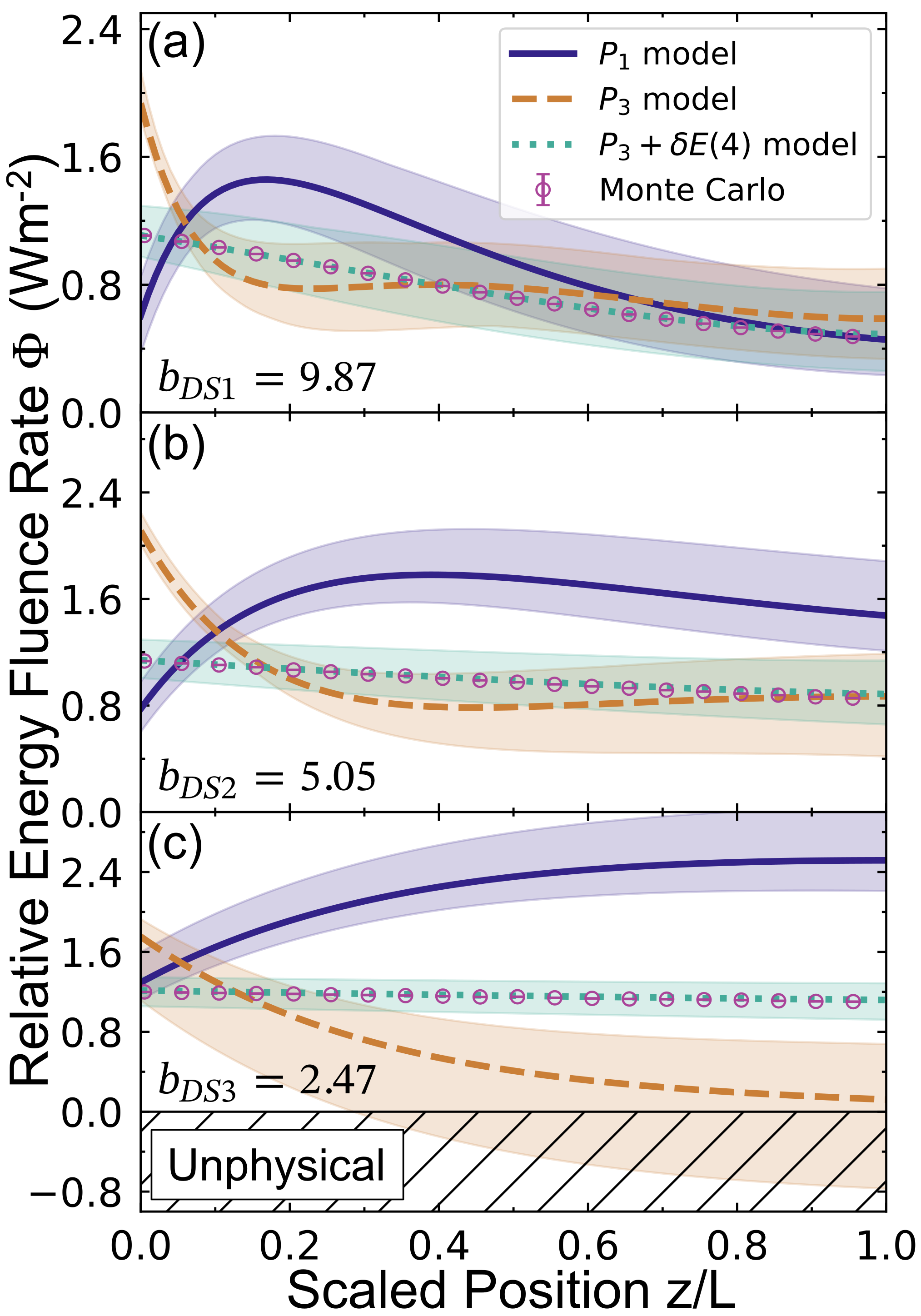}
    \caption{Modeled position-dependent energy fluence rate $\Phi(z)$ for samples (a) DS1, (b) DS2, and (c) DS3.
    The horizontal axis represents the $z$ positions scaled by the thickness of the sample.
    The vertical axis displays the modeled relative $\Phi(z)$.
    Lines represent the results of the $P_{N}$ approximations, and the circles represent the results of Monte Carlo simulations.
    In (c), the unphysical area where $\Phi$ is negative is hatched in black hatched lines.}
    \label{fig:Models_comparison_DS_samples}
\end{figure}

%
Figure~\ref{fig:ModelVsMeasurement_DS_samples} presents the comparison between the models and our measurements.
The $P_1$ and $P_3$ approximations are not shown as they demonstrate poor accuracy (see Figure~\ref{fig:Models_comparison_DS_samples}).
The green areas in Figure~\ref{fig:ModelVsMeasurement_DS_samples} represent the modeled values for the extracted $(a, \: g)$ range using the $P_{3}+\delta E(4)$ approximation.
The horizontal axis represents the $z$ positions scaled by the thickness of the sample, whereas the left ordinate displays the $\Phi(z)$ measurements of samples normalized to the measured the incident intensity, which is presented in Figure~\ref{fig:I0_yScan_and_Fit}(b).
The ordinate corresponds to the modeled relative $\Phi(z)$.
Both ordinates are scaled to compare the trends of measured and modeled $\Phi$, and a good mutual agreement is observed. 
We define a straightforward proportionality, neglecting the small random errors in measurements,
\begin{equation} 
\label{eqn:AssumptionScaling}
	\Phi_{\rm{real}}(z) \equiv K \Phi_{\rm{measured}}(z),
\end{equation}
where $K$ is a constant that represents the portion of $\Phi$ that is measured.
$K$ mainly depends on the experimental limitations that are discussed below in subsection~\ref{subsec:ExperimentalLimitations_3Dsetup}. 
The small differences in measured and modeled $\Phi$ trends, excluding the scaling discussed above, are attributed to the limitations of the Henyey-Greenstein phase function~\cite{Henyey-Greenstein1941APJ}.

The results of the $P_1$, $P_3$, and $P_3 + \rm{\delta E }(4)$ approximations and Monte Carlo simulations are illustrated in Figure~\ref{fig:Models_comparison_DS_samples}, where the $z$ positions are scaled by the thickness of the sample, and the modeled $\Phi(z)$ is relative, as the incident flux density (irradiance) in the models is set to be $F_0 = 1$.
The dynamic ranges of the $P_{N}$ approximations represent the modeled values for the extracted $(a, \: g)$ range. 
Initially, it may appear confusing that a relative result is greater than 1, as seen in the case of sample DS1. 
Although both the irradiance $F_0$ and energy fluence rate $\Phi$ have the same units $\rm{W m^{-2}}$, the former denotes the optical power through the surface of a \textit{flat} unit area in the direction normal to the surface, while the latter refers to the total optical power through a \textit{spherical} unit surface area in all directions~\cite{Henderson2010BookChp}.
Therefore, depending on the scattering properties of the medium under study, it is possible for $\Phi$ to exceed the incident irradiance~\cite{Star1989BookChp, Jacques1998PP, Ong2016OE}. 
%
%

Figure~\ref{fig:Models_comparison_DS_samples} shows that the $P_3 + \rm{\delta E }(4)$ approximation agrees best with the Monte Carlo simulations, consistent with previous reports~\cite{Star1989BookChp, Akdemir2022PRA}.
The $P_1$ and $P_3$ approximations perform poorly for these anisotropically scattering and absorbing samples, and the $P_3$ approximation even predicts an unphysical negative energy fluence rate for the dynamic range of the optically thin sample DS3.
The unphysical results of the $P_3$ approximation for sample DS3 also agree with our previous findings~\cite{Akdemir2022PRA}, which shows a broader unphysical range of $P_3$ for optically thin samples in the strong forward scattering range, compared to the thicker samples.

\subsection{Position-dependent energy fluence rate of undyed microsphere suspensions} \label{subsec:PositionDependentEnergy_3D_UndyedSamples}
%
\begin{figure}
    \centering
    \includegraphics[width=1.0\columnwidth, keepaspectratio]{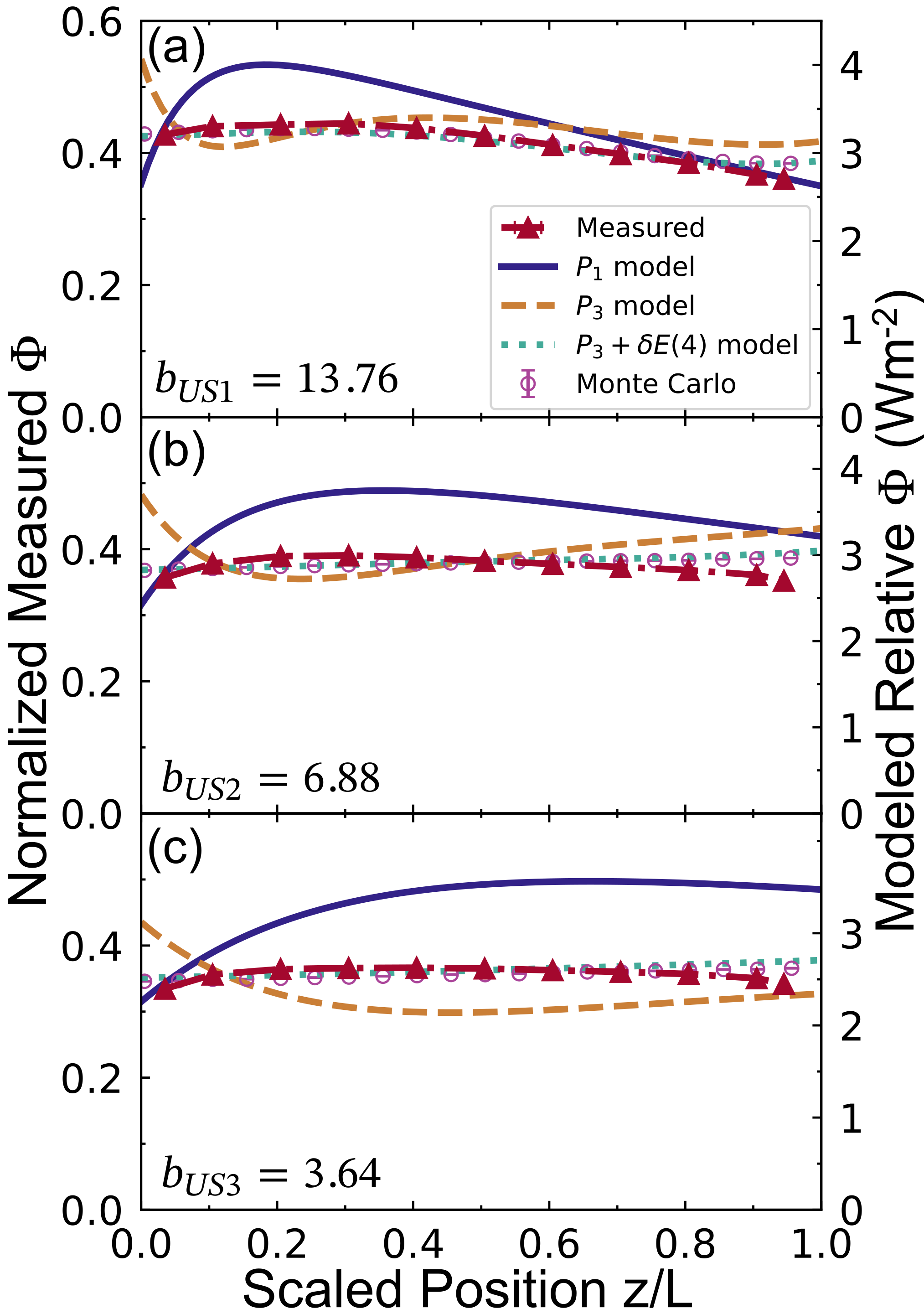}
    \caption{Comparison of observations and models for samples (a) US1, (b) US2, and (c) US3. 
    The abscissa represents the $z$ position scaled by the sample thickness.
    The left ordinate displays the $\Phi(z)$ measurements of samples normalized to the measured incident intensity. 
    The right ordinate corresponds to the modeled relative $\Phi(z)$. 
    The red triangles connected with dash-dotted lines represent the measured results of samples.
    The measured results are calculated by integrating the $y$ scans at each $z$ position, shown as triangles.
    Curves are the results of the $P_{N}$ approximations and the purple circles are the results of Monte Carlo simulations. 
    }
    \label{fig:ModelVsMeasurement_US_samples}
\end{figure}
We present measured $\Phi(z)$ of undyed sphere suspensions and compare them to models in Figure~\ref{fig:ModelVsMeasurement_US_samples}. 
The $P_N$ approximation results do not have dynamic ranges in Figure~\ref{fig:ModelVsMeasurement_US_samples} as the $(a,g)$ pair of samples US1, US2, and US3 are extracted from exact Mie calculations, as explained in subsection~\ref{subsec:Finding_a_g_pair}.
Similar to Figure~\ref{fig:ModelVsMeasurement_DS_samples}, the horizontal axis denotes the $z$ positions scaled by the thickness of the sample, and the left vertical axis gives the $\Phi(z)$ measurements of samples normalized to the measurements of the incident intensity given in Figure~\ref{fig:I0_yScan_and_Fit}(b).
The right vertical axis is the modeled relative $\Phi(z)$, and both vertical axes are scaled to compare the trends of measured and modeled $\Phi$.
A good agreement between the trends of $P_{3} + \delta E(4)$, Monte Carlo, and measurements are observed.
For these microspheres no unphysical predictions occur since the absorption is absent, hence the albedo is equal to 1.
Differences due to experimental limitations are discussed in subsection~\ref{subsec:ExperimentalLimitations_3Dsetup}.

\subsection{Experimental limitations} \label{subsec:ExperimentalLimitations_3Dsetup}
Systematic errors are inevitable in any experiment, including the measurements of $\Phi(z)$. 
It is thus important to identify and comprehend the underlying causes of the systematic errors, before claiming that the measurements reflect reality.
Hence, this subsection discusses the limitations of our experiment that affects the scaling value, $K$, for eqn.~\ref{eqn:AssumptionScaling}.

\begin{figure}
    \centering
    \includegraphics[width=1\columnwidth, keepaspectratio]{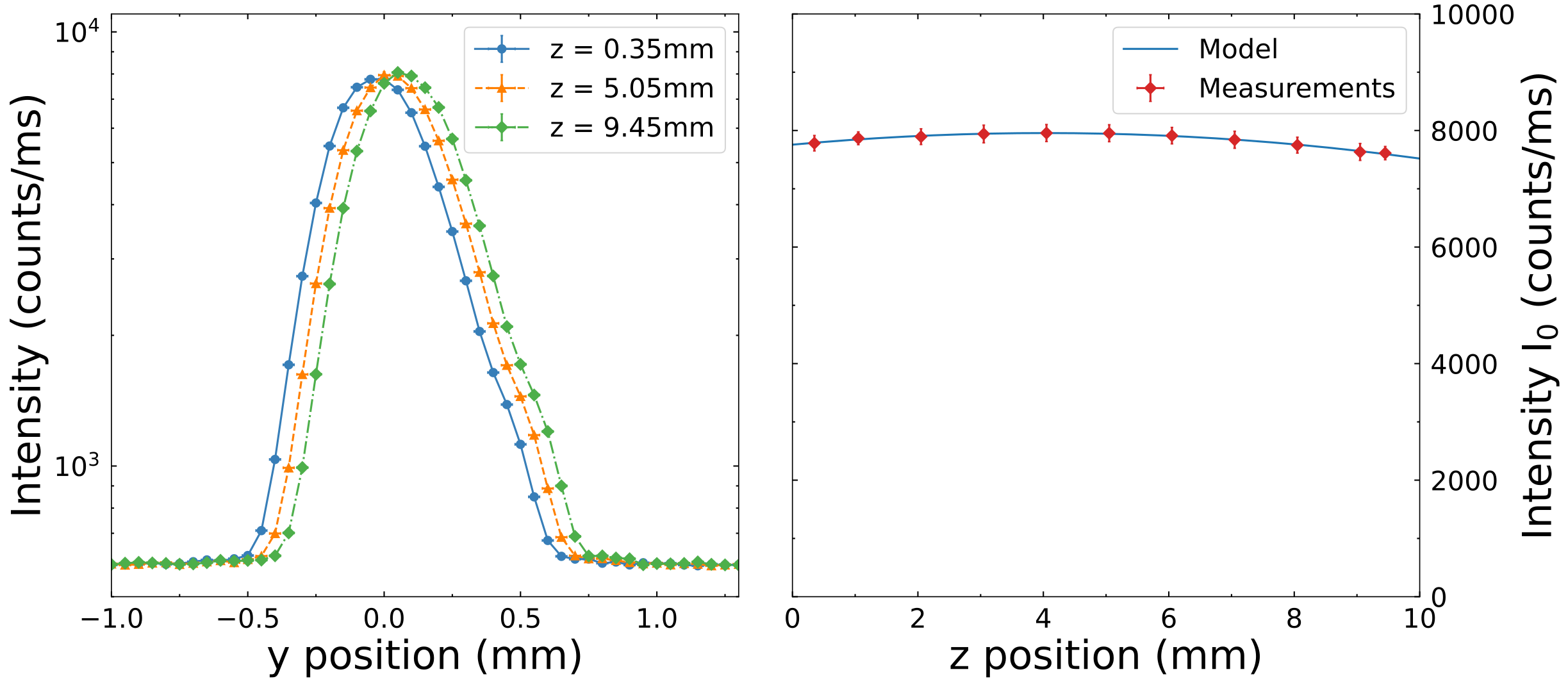}
    \caption{(a) Probe scans in the $y$ direction in absence of a sample.
    This represents the incident beam profiles at the indicated $z$ positions inside the sample.
    (b) Integrated $y$ scans at specific $z$ positions, in the absence of a sample. 
    The blue line represents the polynomial fit to the measurements, to determine the incident beam intensity at all $z$ positions.
    }
    \label{fig:I0_yScan_and_Fit}
\end{figure}

Our discussion starts with the shift in detected peaks for $y$ scans at different $z$ positions, which is not immediately obvious for scans presented in Figure~\ref{fig:Probe_yScan_DS2}.
To clarify this effect, $y$ scans by the probe at specific $z$ are presented in Figure~\ref{fig:I0_yScan_and_Fit}(a), which demonstrate a clear shift of the detected intensity in the $y$ direction. 
This shift is caused by the probe moving at a small angle with respect to the $z$ axis.
The angle is estimated as approximately $\theta_{\rm{z}} = 0.63^{\circ}$, considering the peaks movement of $0.1$ mm in the $y$ direction for a $9.1$ mm travel in the $z$ direction.
Furthermore, Figure~\ref{fig:I0_yScan_and_Fit}(b) presents the incident intensity detected by the probe in air, which is used to normalize the measurements of samples in order to compensate for the effect of the $y$-shift.

%
%

Other experimental limitations associated with the measurements with the probe are listed as follows:
\begin{enumerate}
\item The capillary filled with quantum dots is not exactly parallel to the $x$ axis given in Figure~\ref{fig:SetupSketch}.
Although the kinematic mount attached to the probe holder, as shown in Figure~\ref{fig:ProbeHolder}, is utilized to adjust the capillary angle, perfect alignment is not always achieved.
As a result, the probe positioning inside the sample is restricted to a range between $z=0.35$ mm and $z=9.75$ mm, for the $z$ direction.
In the event that the probe approaches the boundaries beyond these limits, the capillary may come into contact with the quartz walls of the cuvette, potentially causing its displacement or, worse, breaking of the capillary.

\item The stabilities of the probe holder, sample holder, and fiber-optic cable that connects the probe and detector are not perfect.
Small variances in different scan directions presented in Figure~\ref{fig:Probe_zScan_Various_Directions} are likely caused by this.
However, even slight changes in the adjustment of these elements can lead to drastic changes in the detected intensity.
Therefore it is crucial to normalize the measurements to the incident intensity measured under identical conditions as the sample measurements.
The measurements presented in this work are conducted with utmost care to address this issue.

\item The finite dimensions of the cuvette used in our experiments differ from ideal slab geometry, as a slab has an infinite $xy$ plane.
However, the beam diameter of our source ($d_{\rm{beam}} \approx 1$ mm) is significantly smaller than the inner $xy$ dimensions of the cuvette ($42.5 \times 30$ mm), rendering this effect negligible.

\item Although the measurement of $\Phi(z)$ along the $xy$ plane is carried out through $y$ scans of the samples, the scans are limited to the range $y=-6$ mm to $y=6$ mm with the beam centred at $y=0$.
This range is selected due to low signal strength beyond this range, and to shorten the complete measurement time.
The range restriction leaves $18$ mm of unmeasured space in the $y$ axis.
Additionally, during the sample scans, the capillary tip is situated at a distance of approximately $x \approx 11$ $\rm{mm}$ from the cuvette's bottom.
Consequently, the probe scans a rectangular area of dimensions $34 \times 12$ $\rm{mm}$ within the cuvette.
Hence, the portion of the scanned area to the full $xy$ plane within the sample is $A_{\rm{scan}} = 32 \%$.
However, the measurement of $\Phi(z)$ cannot be directly scaled with $A_{\rm{scan}}$ because the experiments employ a collimated beam as the source and detect most of the light within the cuvette, as evident by the low signal strength beyond the scan range (see Figure~\ref{fig:Probe_yScan_DS2}).
However, it is worth noting that the signal outside the scanned range is higher for undyed samples, owing to the lower absorption by the scatterers compared to the dyed samples.
Therefore, the scaling difference between the measured and modeled $\Phi(z)$ for undyed samples is greater than the one for dyed samples.

\item The quantum dots inside the capillary have a certain quantum yield to convert absorbed blue light to emitted red light. 
This quantum yield is provided by the manufacturer as $\eta \geq 64\%$.
Furthermore, the manufacturer specifies the molar extinction coefficient $\epsilon = 5.7 \times 10^{6}$ $\rm{cm^{-1} M^{-1}}$, and the molar concentration $c = 1.0 \pm 0.1$ $\rm{\mu M}$. 
Using the inner diameter of the capillary that holds the quantum dots $d_{\rm{inner}} = 100$ $\rm{\mu m}$ and the Beer-Lambert-Bouguer's law, we calculate the extinction $E_{\rm{qdots}}$ to be
\begin{equation}
	E_{\rm{qdots}} = \epsilon c d_{\rm{inner}} = 57,
\end{equation}
and the optical thickness $b_{\rm{qdots}}$ to be
\begin{equation}
	b_{\rm{qdots}} = E_{\rm{qdots}} * \rm{ln(10)} = 131.2.
\end{equation}

\item The presence of the probe inside the sample has an impact on light transport within the medium.
In order to investigate this effect, the unscattered transmission $T_{\rm{u}}$ of the probe is measured
\footnote{This measurement is conducted with a similar capillary as the one used in the measurements of $\Phi(z)$.}, while scanning in the $y$ direction at various $z$ positions inside sample Ref (see Table~\ref{tab:SampleProperties}).
The obtained results are illustrated in Figure~\ref{fig:ProbeTunscat}.
Based on Figure~\ref{fig:ProbeTunscat}, the extinction coefficient of the probe is derived as $\mu_{\rm{ext}} = 4.8$ $\rm{mm^{-1}}$ when the probe is situated at the center of the beam.
The $\mu_{\rm{ext}}$ represents the sum of absorption and scattering by the probe, hence, the absorbed portion of light by the probe and quantum dots inside, cannot be estimated by this measurement alone.

\begin{figure}
\centering
\includegraphics[width=1.0\columnwidth, keepaspectratio]{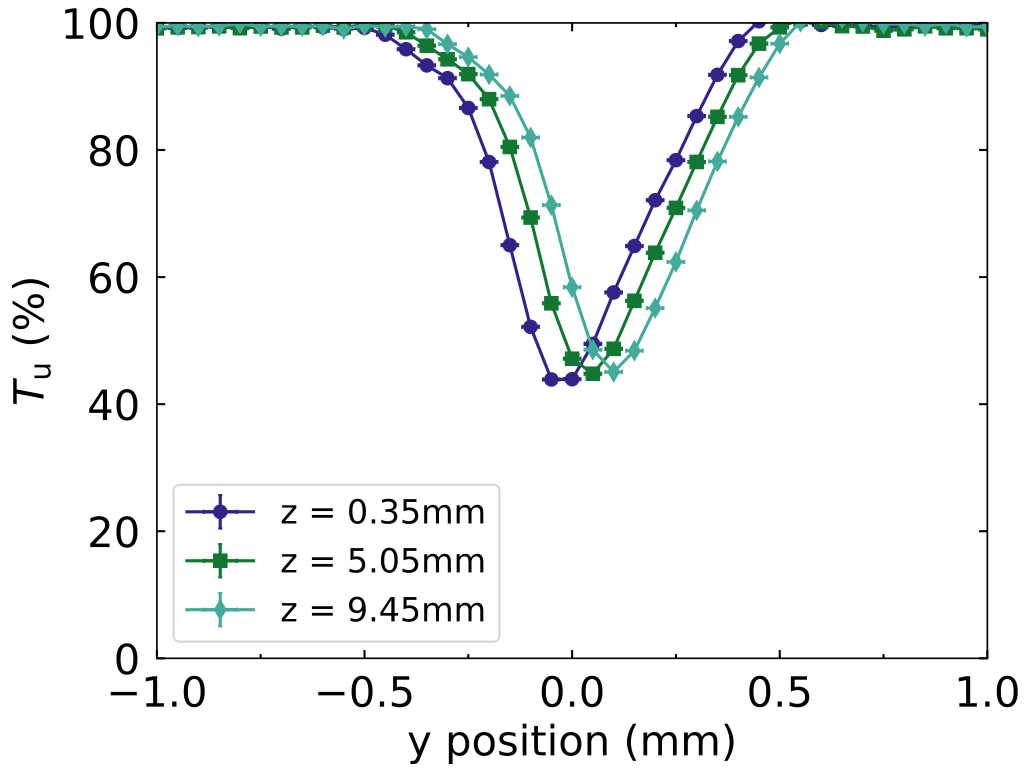}
\caption{Unscattered transmission $T_{\rm{u}}$ of the probe scanning the $y$ direction at various $z$ positions inside sample Ref, see Table~\ref{tab:SampleProperties}.
}
\label{fig:ProbeTunscat}
\end{figure}

\item Part of the emitted light propagates through the capillary and emerges from its top, where it is coupled to the fiber-optic cable of the spectrometer via a connector manufactured in the workshop of the University of Twente. 
This connector, visible in Figure~\ref{fig:ProbeHolder}, introduces a distance of approximately $d_{\rm{cap-fiber}} \approx 5$ mm between the capillary and the fiber.
Assuming that the light emanates from the capillary with a solid angle of $2\pi$ and taking into account the core diameter of the fiber-optic cable ($600$ $\rm{\mu m}$), straightforward calculations yield a coupling efficiency of $\eta_{\rm{coup}} = 2 \%$ between the capillary and the fiber.
\end{enumerate}

All experimental limitations listed above contribute to the deviation of measurement results to the model predictions in scale, and the parameter $K$ in eqn.~\ref{eqn:AssumptionScaling} is introduced to capture these effects.

\section{Conclusions}\label{sec:Conclusion_Probe3D}
In this work, we present a comprehensive analysis of the experimental measurements of the position-dependent energy fluence rate in 3D scattering samples with anisotropically scattering and absorbing microspheres ($r = 0.5$ $\rm{\mu}$m), with and without absorbing dye. 
A thin capillary filled with quantum dots is used as a probe to measure the position-dependent energy fluence rate inside the samples, and the results are compared to analytical models and Monte Carlo simulations. 
The analytical $P_1$ and $P_3$ approximations to the radiative transfer equation predict inaccurate (and even unphysical) results for these samples, that have a range of scatterer densities. The albedo and anisotropy parameters for the dyed samples are extracted from both experiments and Monte Carlo simulations, while Mie calculations are used to determine the parameters for the undyed samples.

Notably, the $P_1$ and $P_3$ approximations fail to match the trends observed in the position-dependent fluence rate, as anticipated. 
In contrast, the Monte Carlo simulations and the $P_3 + \delta E(4)$ approximation, which do not yield unphysical results for the samples under consideration, exhibit a relatively good agreement with the measured trends of the fluence rate.
We provide a detailed discussion of the experimental limitations that prevent an absolute measurement, and an exact match with the models.

\section{Acknowledgments}\label{sec:Acknowledgments_Probe3D}
We thank Cock Harteveld and Melissa Goodwin for their help in sample preparation and building the experimental setup, Mario Vretenar for 3D-printing parts for the experimental setup, and Alwin Kienle, Suhyb Salama, Johannes de Boer, and Gilles Vissenberg for very helpful discussions. 
We thank the staff of the Design Lab and of the Rapid Prototyping Lab at the University of Twente for 3D-printing sample holders for our experimental setup. 
This work was supported by the NWO-TTW program P15-36 ‘Free-form scattering optics’ (FFSO) in collaboration with TUE and TUD and with industrial partners ASML, Demcon, Lumileds, Schott, Signify, and TNO, and by the MESA+ Institute's section Applied Nanophotonics (ANP). 
The data used for this publication are available via the open-access repository Zenodo database that is developed under the European OpenAIRE program and operated by CERN~\cite{ZenodoDoi}. 

\appendix
\section{Infrastructure}
\subsection{Choice of the light source}\label{subsec:ChoiceLightSource}
\begin{figure}[h!]
    \centering
    \includegraphics[width=0.9\columnwidth, keepaspectratio]{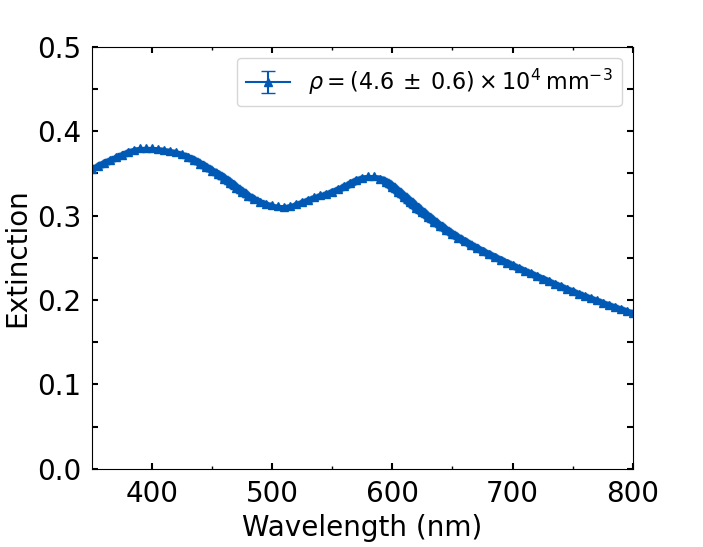}
    \caption{Measured extinction of dyed microsphere suspension with number density $\rho = (4.6 \: \pm \: 0.6) \times 10^4 \: \rm{mm}^{-3}$.
    }
    \label{fig:ExtinctionAnalysis}
\end{figure}
Extinction spectra of the diluted dyed microsphere suspension with a scatterer number density of $\rho = (4.6 \:\pm \: 0.6) \times 10^4 \: \rm{mm}^{-3}$ were measured using a UV-Vis spectrometer (Perkin Elmer Lambda 850, MNF chair at MESA+). 
The measured spectrum is presented in Figure~\ref{fig:ExtinctionAnalysis}. 
The sphere suspension is diluted with DI water, thus, the reference used in this UV-Vis measurement is also DI water. 
The extinction peak of the microspheres in the visible range is observed to be around 400 nm wavelength. 
Therefore, we chose a Thorlabs NPL41B nanosecond laser with a peak wavelength of $\lambda = 408.5 \: \rm{nm}$ as the light source in our experiments to minimize the albedo of the scatterers. 

\subsection{Preparation of the probe} \label{subsec:Probe_3D}
\begin{figure}
    \centering
    \includegraphics[width=1.0\columnwidth]{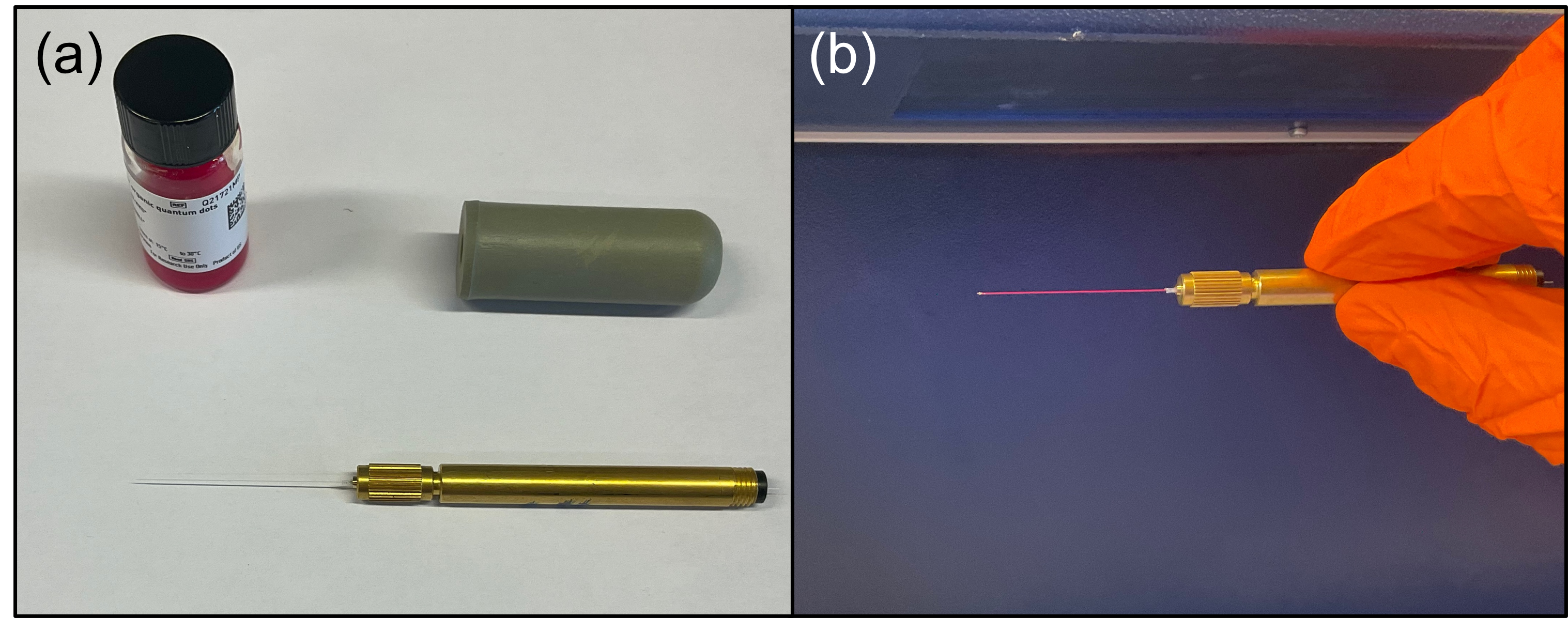}
    \caption{Photograph of the preparation phases of the probe.
    (a) The quantum dots, the capillary held by the fiber chuck, and the rubber stop used for sucking quantum dots inside the capillary.
    (b) Red emission is observed when the capillary is illuminated with UV light.
    }
    \label{fig:ProbePhotos}
\end{figure}

An example of a probe filled with fluorescent reporters is depicted in Figure~\ref{fig:SuspensionSample_TopDetection}. 
To realize such a probe, a cylindrical, quartz capillary with $10 \: \rm{cm}$ length (CM Scientific, CV1017Q-100), an outer diameter of $170\:\rm{\mu m}$, and an inner diameter of $100\:\rm{\mu m}$ is utilized. 
Such a capillary is filled with quantum dots suspended in decane (Thermo Fisher Qdot\textsuperscript{TM} 655 ITK\textsuperscript{TM} Organic Quantum Dots). 
The quantum dots scatter and absorb the incident UV light, and re-emit in the red part of the visible spectrum, with a peak at $\lambda_{\rm{peak}} = 665\: \rm{nm}$ wavelength, see Figure~\ref{fig:ProbePhotos}(b).
To place the quantum dots inside the capillary, the capillary is placed in a teflon-lined fiber chuck (Newport FPH-J). 
A rubber stop is attached to the back end of the fiber chuck, and the quantum dots are sucked in the capillary using this assembly, see Figure~\ref{fig:ProbePhotos}(a). 
Finally, both ends of the capillary are sealed with super glue, as shown in Figure~\ref{fig:ProbePhotos}(c).
In our study, the capillary-fiber chuck assembly is referred to as the \textit{probe}.
%

\subsection{Experimental procedures with the probe}
\label{appx:ProbeProcedure}
\begin{figure}
    \centering
    \includegraphics[width=1.0\columnwidth, keepaspectratio]{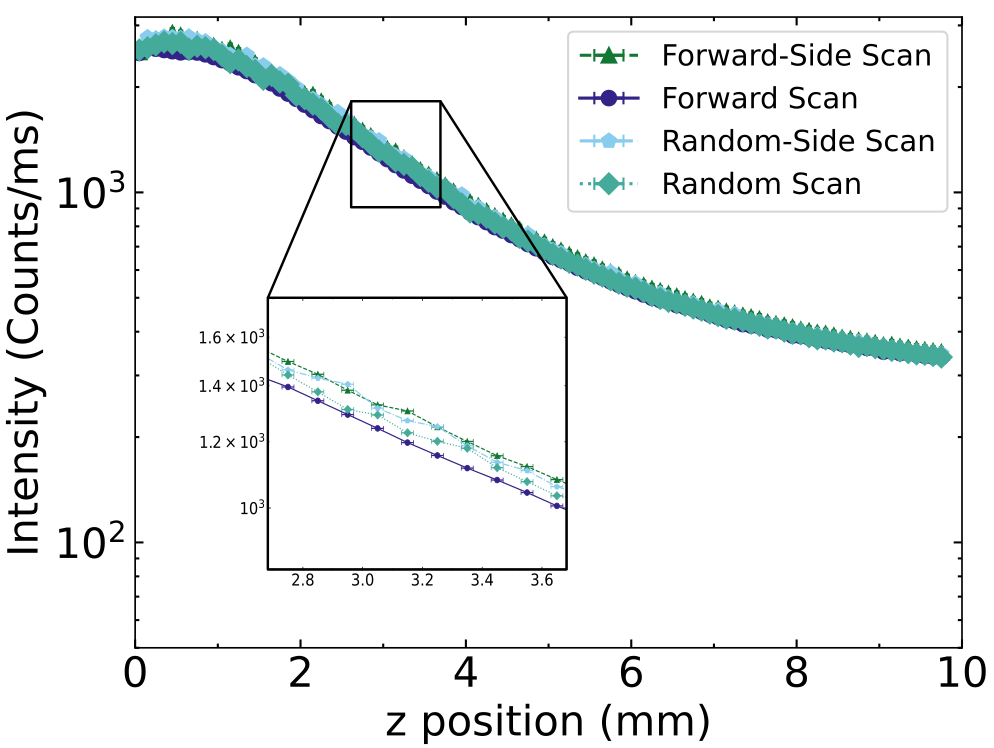}
    \caption{Representative scans of the position-dependent intensity in the $z$ direction in sample DS1, for four different scan directions.
    The vertical axis displays the measured intensity and the horizontal axis shows the position in $z$ direction.
    The inset highlights the small differences between scans.
    }
    \label{fig:Probe_zScan_Various_Directions}
\end{figure}

Measurements with the probe along the $z$ direction inside the sample DS1 are shown in Figure~\ref{fig:Probe_zScan_Various_Directions}, including the results from four primary scan directions: forward, forward-side, random, and random-side. 
To prevent collisions with the sample holder that would damage the fragile probe, the scans start at $z = 0.05$ mm and end at $z = 9.75$ mm. 

The forward direction indicates that the sample is scanned from the left boundary $(z = 0)$ to the right boundary $(z = 10)$, while forward-side refers to scanning the sample in the same direction as forward in $z$, as well as scanning it in the $y$ direction at specific $z$ positions. 
Random direction refers to scans of the sample along the $z$ direction with the probe randomly positioned during the scan.
Finally, random-side indicates that the sample is scanned in the same way as in random direction, but in addition, it is scanned in the $y$ direction at specific $z$ positions, and the $y$ scans are also done with a random positioning of the probe.
Figure~\ref{fig:Probe_zScan_Various_Directions} shows that the scan direction has minimal influence on the outcomes.
The minor differences are attributed to slight instabilities of the probe holder, which are discussed further in subsection~\ref{subsec:ExperimentalLimitations_3Dsetup}.

\begin{figure}[h!]
    \centering
    \includegraphics[width=1\columnwidth, keepaspectratio]{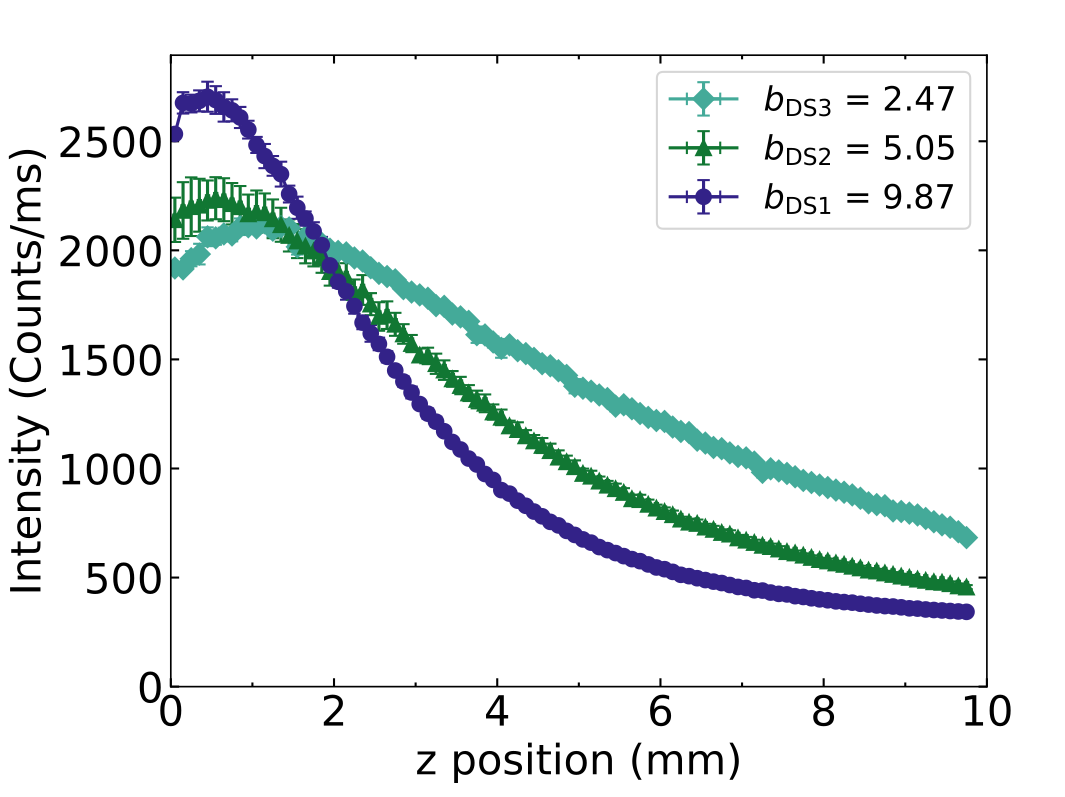}
    \caption{Position-dependent intensity in the $z$ direction detected inside dyed microsphere suspensions DS1, DS2 and DS3.
    The vertical axis displays the measured intensity and the horizontal axis shows the position in $z$ direction.
    The optical thicknesses of the samples are given in the legend.
    Data points represent the averages of four scans in different directions (see Figure~\ref{fig:Probe_zScan_Various_Directions}).
    }
    \label{fig:Probe_zScan_3D}
\end{figure}
Figure~\ref{fig:Probe_zScan_3D} presents the position-dependent intensity in the $z$ direction for samples DS1, DS2, and DS3.
The data points in Figure~\ref{fig:Probe_zScan_3D} represent an average of four different scan directions, as elaborated in Figure~\ref{fig:Probe_zScan_Various_Directions}.
Notably, the optically thicker samples exhibit a higher intensity than the thinner samples, up to a depth of approximately 2 mm inside the sample.
This behavior is attributed to the increased number of scatterers in the thicker samples, which scatter the light into the probe more than the less dense samples at the first few millimeters.
After $z = 2$ $\rm{mm}$ depth, the trend reverses and the densest sample has the lowest detected intensity.
This trend is expected, as the extinction is higher for thicker samples.
However, it should be noted that while the probe scans the $z$ direction entirely in this measurement, it is insufficient to fully characterize the $\Phi (z)$, since a large portion of the $xy$ plane is not measured.
The models simplify the problem to one dimension by utilizing the symmetries within the slab and averaging over the $xy$ plane at each $z$ position~\cite{Akdemir2022PRA}. 
The diameter of our probe is $170$ $\rm{\mu m}$, while the beam size is approximately 1 mm, and the $z$ scans are done when the probe is at the center of the beam in $y$ direction. 
Consequently, although the probe's length in the $x$ direction is roughly 40 mm within the sample, the scan presented in Figure~\ref{fig:Probe_zScan_3D} predominantly collects the unscattered intensity.
%
\begin{figure}[h!]
    \centering
    \includegraphics[width=1\columnwidth, keepaspectratio]{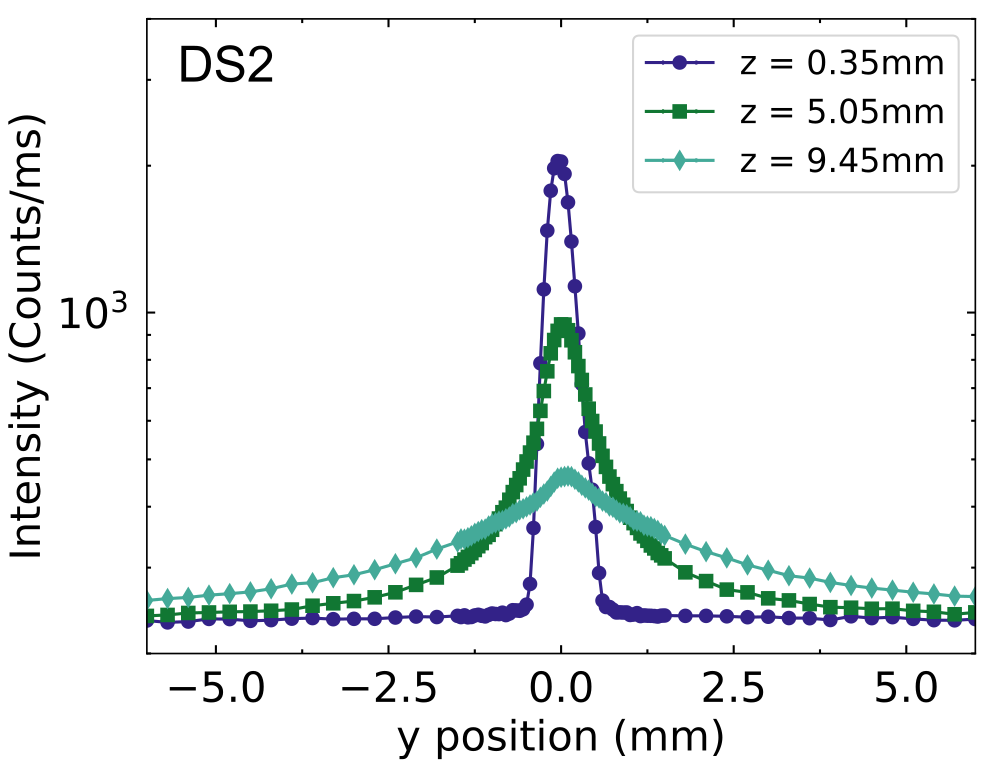}
    \caption{Representative scans of $y$ direction at specific $z$ positions inside sample DS2.
    The vertical axis displays the measured intensity and the horizontal axis shows the position in $z$ direction.
    }
    \label{fig:Probe_yScan_DS2}
\end{figure}

In order to obtain a more comprehensive understanding of $\Phi(z)$, it is necessary to scan samples in the $y$ direction at various positions in $z$.
Figure~\ref{fig:Probe_yScan_DS2} presents three scans at specific $z$ positions.
The results show that the energy distribution broadens in the $y$ direction as the light travels deeper in $z$, in agreement with expectations from multiple scattering~\cite{Koenderink2005JOSAB, Bret2005Thesis, Akkermans2007Book, Carminati2021book}.
All samples are probed in the $y$ direction at numerous $z$ positions, and the resulting curves obtained from these $y$ scans at each $z$ position are integrated to derive a value that represents the total $\Phi(z)$ integrated over an $xy$ plane.
\begin{figure}[h!]
    \centering
    \includegraphics[width=0.9\columnwidth, keepaspectratio]{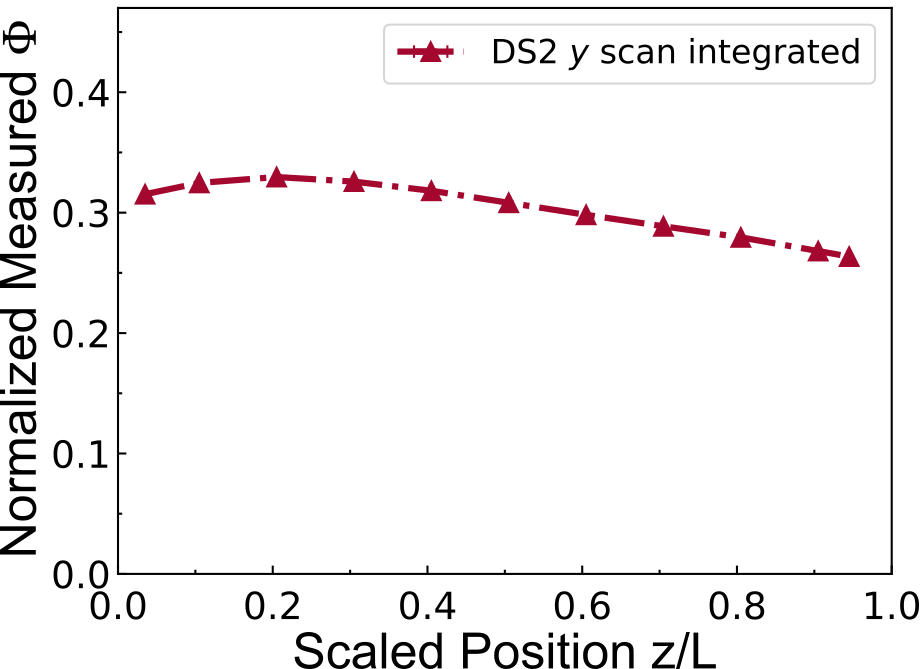}
    \caption{Representative integrated $y$ scans at specific $z$ positions inside sample DS2.
    The vertical axis displays the $\Phi(z)$ measurements of samples normalized to the measurements of the incident intensity.
    The horizontal axis represents the $z$ positions scaled by the thickness of the sample.
    }
    \label{fig:Probe_yScan_integ_DS2}
\end{figure}
Figure~\ref{fig:Probe_yScan_integ_DS2} shows the integrated $y$ scans of sample DS2, where the $z$ positions are scaled by the thickness of the sample, and the measured $\Phi (z)$ is normalized by the incident intensity.
Notably, Figure~\ref{fig:Probe_yScan_integ_DS2} illustrates the inclusion of diffuse light to the detection, in contrast to the measurements obtained solely from $z$ scans, as shown in Figure~\ref{fig:Probe_zScan_3D}.
%

\section{Theory background}\label{appx:TheoryBackground}
Transport theory describes the propagation of light in complex media using the well-known radiative transfer equation (RTE)~\cite{Chandrasekhar1960Book, Tait1964Book, Ishimaru1978book, Wang2014PRA, Vynck2016PRA, Arruda2018PRA, Tommasi2020PRA}. 
Whereas in transport theory, one neglects the wave nature and concomitant interference of light (speckle), it is widely used as it generally provides a robust description~\cite{vanRossum1999RMP, Carminati2021book}. 
The most fundamental quantity described by RTE is the specific intensity $I(\mathbf{r}, \hat{s})$ that describes the average power flux density at position $\mathbf{r}$ in a given direction $\hat{s}$, within a unit solid angle and a unit frequency band. 
It is challenging to solve $I(\mathbf{r}, \hat{s})$ directly, as it depends on both the position $\mathbf{r}$ and the direction $\hat{s}$.
Therefore, analytical approximations to RTE or Monte Carlo simulations are preferred.

\subsection{$P_N$ approximations}\label{subsec:PNApprox}
The so-called $P_N$ approximation is a widely used analytical approximation to the RTE, where the dependence on both variables $(\mathbf{r}, \hat{s})$ is separated~\cite{Tait1964Book} by expanding $I(\mathbf{r,\hat{s}})$ in products of complete sets on the domains of $\mathbf{r} $ and $\hat{s}$
%
\begin{equation}{\label{eqn:specIntExp}}
 I(\mathbf{r,\hat{s}}) = \sum_{l = 0}^{N} \sum_{m = -l}^{l}
 \psi_{l}^{m}(\mathbf{r}) Y_{l}^{m} (\hat{s}).
\end{equation}
%
Here, $\psi_{l}^{m}(\mathbf{r})$ are the spatial components, $Y_{l}^{m} (\hat{s})$ are the directional components taken as the Laplace spherical harmonics~\cite{Jackson1998Book,Arfken2005Book}, and $N$ is the order of the approximation that determines the number of terms in Eq.~(\ref{eqn:specIntExp}). 
In particular the $P_1$ approximation is well known as it corresponds to the widely-used diffusion approximation~\cite{vanRossum1999RMP, Carminati2021book}. 
The analytical $P_N$ approximations are mostly used for simple sample geometries such as a slab and a sphere and their accuracies depend notably on (\textit{i}) the order $N$ of the approximation, (\textit{ii})  on the optical properties of the medium and (\textit{iii}) the scattering phase function used in the approximation.

\subsection{Boundary conditions}\label{subsec:BoundaryCond}
In this work, we apply Marshak type boundary conditions~\cite{Modest2003book} to solve the RTE for a slab geometry.
The boundary conditions assume that on the boundaries, no diffuse light enters the scattering medium from the surrounding medium outside.
When the refractive indices of the media inside and outside do not match, the total diffuse flux at the boundary directed into the medium is the part of the outwardly directed flux that is reflected by the surface \cite{Keijzer1988AO, Star1988PMB, Star1989BookChp, Liemert2014MP, Liemert2021JOSAA}.
This conservation law results in two boundary conditions
\begin{equation} \label{eqn:BC1}
    \int_0^1 I_{\rm{diff}}(0,\mu) P_l(\mu) d\mu = \int_0^1 R(\mu) I_{\rm{diff}}(0,-\mu) P_l(\mu) d\mu,
\end{equation}
\begin{equation} \label{eqn:BC2}
    \int_{-1}^0 I_{\rm{diff}}(L,-\mu) R(-\mu) P_l(\mu) d\mu = \int_{-1}^0  I_{\rm{diff}}(L,\mu) P_l(\mu) d\mu,
\end{equation}
where $L$ is the sample thickness, $R(\mu)$ the specular Fresnel reflection function for unpolarized light, $I_{\rm{diff}}$ the diffuse specific intensity, and $\mu$ the directional cosine of light. 
$P_l$ is a Legendre polynomial and $l$ specifies the order of the approximation.

In our experiments, the microsphere suspensions are held inside quartz cuvettes during measurements.
We modify the boundary conditions in eqns.~\ref{eqn:BC1} and \ref{eqn:BC2} by introducing a reflection coefficient that accounts for the internal reflections from the quartz walls surrounding the samples.
The reader is referred to the underlying PhD thesis~\cite{Akdemir2023Thesis} for details about this modification and the full derivation of the $P_N$ approximations used in this work.

\bibliographystyle{apsrev4-1}
\bibliography{_ProbeFluence_3D.bib}

\end{document}